\documentclass[12pt]{article}
\usepackage{bm}
\usepackage{a4}
\usepackage{amsthm}
\usepackage{amsmath}
\usepackage{amssymb}
\usepackage{graphicx}
\usepackage{xcolor}
\usepackage{graphics}
\evensidemargin \oddsidemargin
\definecolor{drab}{rgb}{0.59, 0.44, 0.09}
\definecolor{applegreen}{rgb}{0.55, 0.71, 0.0}
\definecolor{auburn}{rgb}{0.43, 0.21, 0.1}
\definecolor{awesome}{rgb}{1.0, 0.13, 0.32}
\definecolor{britishracinggreen}{rgb}{0.0, 0.26, 0.15}
\definecolor{darksalmon}{rgb}{0.91, 0.59, 0.48}
\definecolor{desert}{rgb}{0.76, 0.6, 0.42}
\def\b#1{{\mathbb #1}}
\newcommand{\CC}{\mathbb{C}}
\newcommand{\RR}{\mathbb{R}}
\newcommand{\ZZ}{\mathbb{Z}}
\newcommand{\NN}{\mathbb{N}}

\def\b{\mathfrak{b}}
\newcommand{\cross}{\mbox{$\rule{0.7pt}{1.3ex}\!\times $}}

\newcommand{\Hi}{{\cal H}}

\newcommand{\A}{{\cal A}}
\newcommand{\B}{{\cal B}}

\newcommand{\bx}{\bm{x}}
\newcommand{\bL}{\bm{L}}
\newcommand{\bpi}{{\bm\pi}}
\newcommand{\bphi}{{\bm\phi}}
\newcommand{\bpsi}{{\bm\psi}}
\newcommand{\bomega}{{\bm\omega}}
\newcommand{\mathsym}[1]{{}}
\newcommand{\unicode}[1]{{}}

\newcommand{\be}{\begin{equation}}
\newcommand{\ee}{\end{equation}}
\newcommand{\bea}{\begin{eqnarray}}
\newcommand{\eea}{\end{eqnarray}}
\newcommand{\ba}{\begin{array}}
\newcommand{\ea}{\end{array}}
\def\nn{\nonumber \\}

\newtheorem{teorema}{Theorem}[section]

\newtheorem{lemma}{Lemma}[section]

\newtheorem{propo}{Proposition}[section]

\begin{document}
\title{
On localized and coherent states on some new fuzzy spheres}
\date{}

\author{Gaetano Fiore, Francesco Pisacane
   \\   \\  
Dip. di Matematica e applicazioni, Universit\`a di Napoli ``Federico II'',\\
\& INFN, Sezione di Napoli, \\
Complesso Universitario  M. S. Angelo, Via Cintia, 80126 Napoli, Italy}

\maketitle

\begin{abstract}
\noindent
We construct various systems of coherent states  (SCS) on the
$O(D)$-equivariant fuzzy spheres $S^d_\Lambda$ ($d=1,2$, $D=d\!+\!1$)  
constructed in 
[G. Fiore, F. Pisacane, 
J. Geom. Phys. 132 (2018), 423-451]
and  study their localizations in configuration space as well as angular momentum space.
These localizations are best expressed through the $O(D)$-invariant  square space and angular momentum 
uncertainties $(\Delta\bx)^2,(\Delta\bL)^2$ in the ambient Euclidean space $\RR^D$.
We also determine general bounds (e.g. uncertainty relations  from commutation relations) for  $(\Delta\bx)^2,(\Delta\bL)^2$, and partly investigate which SCS may saturate these bounds. 
In particular, we  determine  $O(D)$-equivariant systems of optimally localized coherent states,
which are the closest quantum states to the classical states (i.e. points) of $S^d$.
We compare the results with their analogs on commutative $S^d$. We also show that  on $S^2_\Lambda$ our optimally localized states   are better localized than those on the Madore-Hoppe  fuzzy sphere with the same cutoff $\Lambda$.
\end{abstract}

\section{Introduction}
\label{introduzione}

The present interest in noncommutative space(time) algebras  has various motivations.
In particular,  such algebras may describe spacetimes at microscopic scales, 
regularizing ultraviolet divergencies in quantum field theory (QFT) and/or allowing the quantization of gravity, or may help to unify  fundamental interactions (see e.g. \cite{DopFreRob95,GroKliPre96',AlessioArzano,ConLot91,AscMadManSteZou}). 
Noncommutative geometry   \cite{Connes,Madore99,GraFigVar00,Lan97} 
develops the needed machinery of differential geometry  on such algebras. 
Fuzzy spaces are special examples of noncommutative spaces: a fuzzy space is a sequence $\{\mathcal{A}_n\}_{n\in\NN}$ of finite-dimensional algebras such that as
$n$ diverges  $\mathcal{A}_n$ goes to the commutative algebra $\A$
of regular functions on an ordinary manifold. 
The first and seminal fuzzy space is the socalled Fuzzy 2-Sphere (FS) of Madore and Hoppe \cite{Madore,HopdeWNic}: 
${\cal A}_n\simeq M_n(\CC)$ (the algebra of complex $n\times n$ matrices)
is generated by coordinate operators $\left\{x_i\right\}_{i=1}^{3}$ fulfilling
\be
[x_i,x_j]=\frac {2i}{\sqrt{n^2\!-\!1}}\varepsilon^{ijk}x_k, \qquad\qquad
x_ix_i=1                     \label{FS}
\ee
($n\!>\!1$, sum over repeated indices is understood);
in fact these are obtained by the rescaling 
\be
x_i=\frac{2L_i}{\sqrt{n^2\!-\!1}},\quad i=1,2,3                   \label{rescale}
\ee
 of the elements $L_i$ of  the standard basis of $so(3)$ in the 
 {\it irreducible} representation $(\pi_l,V_l)$ characterized by $\bL^2:=L_iL_i=l(l\!+\!1)$,  or equivalently of dimension $n=2l\!+\!1$. 

On the contrary, the Hilbert space ${\cal L}^2(S^2)$  of a quantum particle on $S^2$
 decomposes as the direct sum of {\it all} the irreducible representations of $SO(3)$,
\be
{\cal L}^2(S^2)=\bigoplus\limits_{l=0}^\infty V_l;
\label{directsum}
\ee
the angular momentum components \ $L_i$ \ map the generic $V_l$ into itself, while the $x_i$ map it into $V_{l-1}\oplus V_{l+1}$. 
Moreover, relations (\ref{FS}) are equivariant under $SO(3)$, but not under the whole $O(3)$,  in particular not under parity $x_i\mapsto -x_i$; whereas the commutators of the cooordinates $x_i$ on the sphere $S^2$ remain zero under all $O(3)$ transformations.

In \cite{FiorePisacaneJGP18} we have introduced some new fuzzy approximations 
of $S^1,S^2$  - more precisely a fully 
$O(2)$-equivariant fuzzy circle $\{S^1_\Lambda\}_{\Lambda\in\NN}$ and  a fully  $O(3)$-equivariant fuzzy 2-sphere $\{S^2_\Lambda\}_{\Lambda\in\NN}$; \ the latter
is free of the mentioned shortcomings.
To construct  $S^d_\Lambda$ ($d\!=\!1,2$) we have first
projected the Hilbert space $\mathcal{L}^2(\RR^D)$ of a zero-spin quantum particle  in $\RR^D$ ($D\!:=\!d\!+\!1$) onto the finite-dimensional  subspace $\mathcal{H}_{\Lambda}$ spanned by all the $\psi
$ fulfilling
\be
H_{\Lambda}\psi=E\psi, \qquad \mbox{where }\quad H_{\Lambda}:=-\frac 12\Delta + U_{\Lambda}(r)
, \quad   E\leq\overline{E}(\Lambda):=\Lambda(\Lambda\!+\!d\!-\!1). \label{Heigen}
\ee
Here $r^2:=\bx^2= x_ix_i$, \ $x_i$ ($i=1,...,D$) are  Cartesian coordinates of $\RR^D$
(we use dimensionless variables), \
$\Delta:= \frac{\partial}{\partial x_i}\frac{\partial}{\partial x_i}$ is the Laplacian on  
$\RR^D$,
$U_{\Lambda}(r)$ is a confining potential  with a very sharp minimum at  $r=1$, i.e. 
with $U_{\Lambda}'(1)=0$ and very large $k(\Lambda):= U_{\Lambda}''(1)/4>0$; 
 \  we have fixed $U_{\Lambda}(1)$  so that the lowest energy (i.e. eigenvalue of the Hamiltonian $H_{\Lambda}$) is $E_0=0$. \
In other words,  the subspace $\mathcal{H}_{\Lambda}\subset\mathcal{L}^2(\RR^D)$ 
is characterized by energies below the cutoff $\overline{E}$. 
Passing to the radial coordinate $r$ and angular ones, 
(\ref{Heigen}) is reduced to a 1-dimensional Schr\"odinger equation in an
unknown $f(r)$. This  is well approximated by that of a harmonic oscillator
by further requiring that $U_{\Lambda}$ satisfies the conditions
\be
U_{\Lambda}(r)\simeq U_{\Lambda}(1)+2k (r-1)^2,\qquad
\qquad k(\Lambda)\ge\Lambda^2(\Lambda\!+\!1)^2                     \label{kineq}
\ee
(this guarantees in particular that the classically allowed region 
$U_{\Lambda}(r)\le \overline{E}$ is a thin spherical shell of radius $\simeq 1$); 
by the second we also exclude 
all radial excitations from the part of the spectrum of $H_{\Lambda}$  below $\overline{E}$
and make the latter coincide [up to terms $O(1/\Lambda)$ depending on 
higher order terms in the Taylor expansion of $V_\Lambda$] with that 
of the Hamiltonian of free motions (the Laplacian) on $S^d$,  $\bL^2:=L_{ij}L_{ij}/2$; 
\ here $L_{ij}:=i(x_j \frac{\partial}{\partial x_i}\!-\!x_i \frac{\partial}{\partial x_j})$ are the angular momentum components. Denoting as $P_{\Lambda}$ the projection on
$\mathcal{H}_{\Lambda}$, to every observable $A$ on $\mathcal{L}^2(\RR^D)$  we can associate  one $\overline{A}:=P_{\Lambda} A P_{\Lambda}$  on 
$\mathcal{H}_{\Lambda}$.
In particular we have computed at leading order in $1/\Lambda$ the action of 
 $\overline{x_i}$, $\overline{L_{ij}}$ on
$\mathcal{H}_{\Lambda}$ and the algebraic relations that they fulfill. We
have fine-tuned the definition of the ``fuzzy'' Cartesian coordinates 
$\overline{x_i}$ and angular momentum components  $\overline{L_{ij}}$ in the 
simplest way, allowed by the residual freedom of choice of $V_\Lambda$;
these relations are reported at the beginnings of sections  \ref{CSD=2}, \ref{CSD=3}.
The resulting algebra  
$\mathcal{A}_{\Lambda}=End(\mathcal{H}_{\Lambda})$ of fuzzy observables is 
equivariant under the full group $O(D)$ of orthogonal transformations (including inversions of the axes), is generated by the $\overline{x_i}$ and is spanned by
ordered monomials in $\overline{x_i}$, $\overline{L_{ij}}$. \ 
In particular \ $[\overline{x_i},\overline{x_j}]$ \  is proportional to $\overline{L_{ij}}$, as in  Snyder noncommutative space\footnote{Snyder's quantized spacetime algebra is generated by 4 hermitean Cartesian coordinate operators $\left\{x^\mu\right\}_{\mu=0,1,2,3}$, 
 and 4 hermitean momentum operators $\left\{p_\mu\right\}_{\mu=0,1,2,3}$  fulfilling (here $\alpha$ is a suitable constant)
\begin{equation}
[p_\mu,p_\nu]=0, \qquad
[x^\mu,p_\nu]=i\hbar(\delta^\mu_\nu-\alpha p^\mu p_\nu), \qquad [x^\mu,x^\nu]=-i\hbar\alpha L^{\mu\nu},\qquad\qquad
\mu,\nu=0,1,2,3
\end{equation}
where  $L^{\mu\nu}
=x^\mu p^\nu-x^\nu p^\mu$ and
$v^\mu=\eta^{\mu\nu}v_\nu$, with $\eta=\mbox{diag}(1,-1,-1,-1)=\eta^{-1}$
the Minkowski metric matrix.
} 
\cite{Snyder} and in some higher dimensional fuzzy spheres \cite{GroKliPre96,Ramgoolam,Dolan:2003kq,Ste17}. 
Actually, there is an $O(D)$-equivariant realization of $\mathcal{A}_{\Lambda}$
in terms of an irreducible vector representation of $Uso(D+1)$: the
$\overline{L_{ij}}$ are realized exactly  as the elements \ $L_{ij}\!\in\!so(D) \!\subset\! so(D+1)$, \ while
the $\overline{x_i}$ are realized  as the elements \ $L_{(D+1)i}\!\in\! so(D+1)$ \ multiplied by factors depending only on $\bL^2$. 
Below we shall remove the bar and denote $\overline{x_i}$, $\overline{L_{ij}}$
again as $x_i,L_{ij}$.  Moreover,
the Hilbert space $\mathcal{H}_{\Lambda}$ on  $S^2_\Lambda$ decomposes as  the direct sum \ $\mathcal{H}_{\Lambda}=\bigoplus\nolimits_{l=0}^\Lambda V_l$; \ 
the angular momentum components \ $L_i=\varepsilon^{ijk}L_{jk}/2$ \ map the generic  $V_l$ into itself, while the coordinates $x_i$ map it into $V_{l-1}\oplus V_{l+1}$, as
they do on ${\cal L}^2(S^2)=\bigoplus_{l=0}^\infty V_l$. For these reasons we believe
that in the $\Lambda\!\to\!\infty$ limit  the fuzzy sphere $S^2_\Lambda$  approximates the configuration space $S^2$ better than the FS does. 

As known \cite{Schroe26,Klauder,Glauber,PerelomovCMP72}, systems of coherent states (SCS) are an extremely useful tool for studying quantum theories with both finitely and infinitely many degrees of freedom (QFT). In particular, 
they may decisively simplify the computation of path integrals representing propagators, correlation functions and their generating functionals; this is applied in  nuclear, atomic, condensed matter and elementary particle physics (see e.g. \cite{Klauder-Skager85,AliAntGaz13,AntBagGaz18,FioGueMaiMaz15}).  From a foundation-minded viewpoint, Berezin's
quantization procedure   on  K\"ahler manifolds   \cite{Berezin1,Berezin2,Berezin3} itself is based on the existence of SCS.
For the same reasons  the search for coherent states is crucial \cite{BarrettEtAl2011,Ste16NPB} also  for quantum theories on fuzzy manifolds (see e.g. \cite{GroMad92,GroKliPre96',GroKliPre96,AscMadManSteZou,Medina:2002pc}).
Standard SCS $\{|\alpha\rangle\}_{\alpha\in\CC}$ on the phase plane  can be defined in 3 equivalent ways: 

 (A) $|\alpha\rangle$ saturates Heisenberg uncertainty relation  (HUR).

(B) $|\alpha\rangle$ is an eigenstate of the annihilation operator with eigenvalue $\alpha\in\CC$.

(C) $|\alpha\rangle$ is generated by the Heisenberg-Weyl group operator $D(\alpha)$ acting on  vacuum $|0\rangle$. 

\noindent
Perelomov \cite{PerelomovCMP72,Perelomov} defines generalized CS on orbits of various Lie groups $G$
basically using (C); $|0\rangle$ can be any vector in the carrier Hilbert space. 
If $|0\rangle$ maximizes the  isotropy subalgebra $\b$
in the complex hull of the Lie algebra of $G$, then it  is also annihilated by some element(s) in $\b$,  the corresponding CS are eigenvectors of the latter
and minimize a specific $G$-invariant uncertainty [$(\Delta \bm{L})^2$  in the case $G=SO(D)$, see
below];  in this sense also properties (A), (B) are fulfilled.

The main aim of the present work is to introduce  on $S^d_\Lambda$ ($d=1,2$) various systems of coherent states  (SCS). We follow in spirit Perelomov's approach, with $G$ the isometry group $O(D)$ of $S^d$. However, our Hilbert space 
$\mathcal{H}_{\Lambda}$
will in general carry a {\it reducible} representation  of $O(D)$; moreover,
we study the localization properties of these SCS also in configuration space,
beside in (angular) momentum space.
We consider SCS both in the {\it strong sense},
i.e. providing a resolution of the identity, and in the {\it weak sense}, i.e. making up
just an (over)complete set in $\Hi_{\Lambda}$. On $\mathcal{H}_{\Lambda}$ the uncertainties $\Delta x_i,\Delta L_{ij}$  must fulfill a number of  uncertainty
relations and other
 inequalities following from the algebraic relations (commutation, etc.) among the 
$x_i,L_{ij}$.  Neither on the commutative nor on the fuzzy spheres is it possible to saturate all of them (and their consequences, a fortiori). 
Therefore we preliminarly discuss the saturation of  suitable $O(D)$-invariant inequalities first on $S^d$,  then on $S^d_\Lambda$ , because they have a physical meaning independent of the particular
chosen reference frame, and because a state saturating them is automatically mapped
into another one by the unitary transformation $U(g)$ corresponding to any
orthogonal transformation $g\in O(D)$ \ [by definition $g_{ij}x_j=U^{-1} (g)\, x_i\, U(g)$, etc.]. 
\begin{figure}
\begin{center}
\includegraphics[height=8cm]{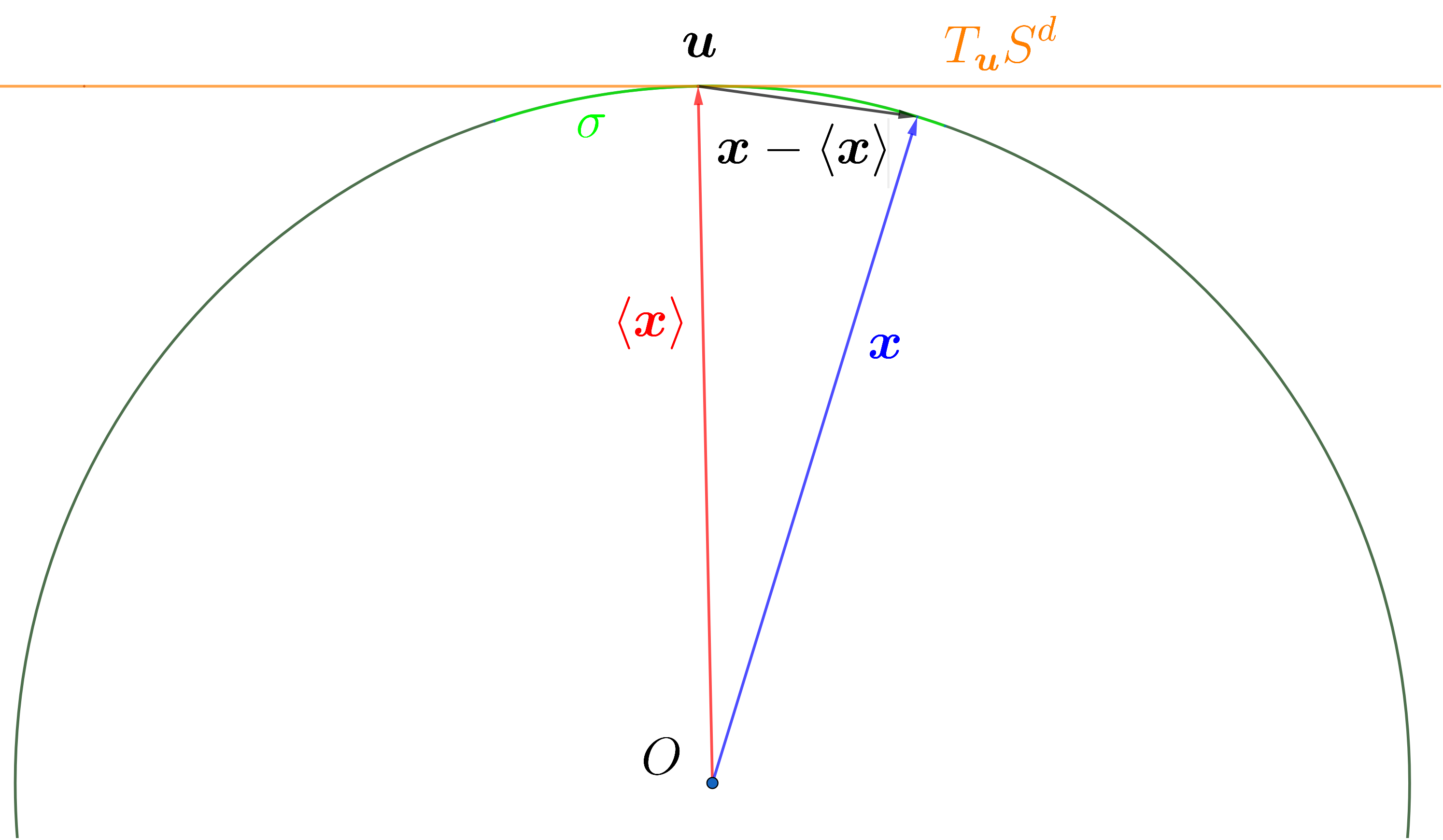}
\end{center}
\caption{The vectors {\color{blue}$\bm{x}$}, {\color{red}$\left\langle\bm{x}\right\rangle$}, $\bm{x}-\left\langle\bm{x}\right\rangle$, the region {\color{green}$\sigma$} and the tangent plane {\color{orange}$T_{\bm{u}}S^d$} at $\bm{u}$.}
\label{Vett_tg}
\end{figure}
More precisely, as a measure of localization of a state  in configuration space we adopt its {\it spacial dispersion}, i.e.
 the expectation value 
\be\label{uncnostra}
(\Delta \bm{x})^2: =\sum_{i=1}^D(\Delta x_i)^2\equiv
\left\langle \left( \bm{x}\!-\!\left\langle\bm{x}\right\rangle\right) ^2\right\rangle
=\left\langle\bm{x}\,{}^2\right\rangle- \left\langle \bm{x}\right\rangle^2
\ee
on the state; here $\bm{x}\equiv(x_1,...,x_n)$, 
$\left\langle \bm{x}\right\rangle\equiv(\left\langle x_1\right\rangle,...,\left\langle x_n\right\rangle)$ pinpoints the average position of the particle  in the ambient Euclidean space  $\RR^D$,
the scalar observable \  $\bx^2:=\sum_{i=1}^Dx_ix_i$ \
measures the square distance from the origin, 
the vector observable $\bm{x}\!-\!\left\langle\bm{x}\right\rangle$ 
measures the displacement from the average position, and expression (\ref{uncnostra}) is the average of the square of the latter.
To motivate this choice we note that it is manifestly $O(D)$-invariant and that
if the state is localized in a small region $\sigma\subset S^d$
around a point $\bm{u}\equiv\left\langle\bm{x}\right\rangle\in S^d$ then
$(\Delta \bm{x})^2$ essentially reduces to the average square displacement
in the tangent  plane at $\bm{u}$, see fig. \ref{Vett_tg}: the
metric on the sphere is induced by the one in the ambient Euclidean space, as wished.
Eq. (\ref{uncnostra}) can be seen as a generalization of the  square dispersion $(\Delta\bm{L})^2$ of  the spin $\bm{L}$ as introduced by Perelomov \cite{Perelomov}, to which it reduces 
upon replacing $\bm{x}$ by $\bm{L}$. In fact,  as a measure of localization of the state  in (angular) momentum space we shall adopt $(\Delta\bm{L})^2$.

Given a state, consider an orthogonal transformation $g\!\in\! O(D)$ such that $g\!\left\langle \bx \right\rangle\!=\!(|\langle \bm{x}\rangle|,\!0,\!...,\!0)$; then the state  is mapped  by $U(g)$ into a new one with the same \ $\left\langle \bx^2\right\rangle$, \ $\left\langle x_1\right\rangle=|\!\left\langle \bm{x}\right\rangle\!|$, \ $\left\langle x_i\right\rangle=0$ for $i\!>\!1$  \ (of course one obtains the same result replacing $x_1$ by any other $x_i$, or by the $L_i$). If $\bx^2$ is central in the algebra of observables and the representation of the latter is irreducible, then $\left\langle \bx^2\right\rangle$
is state-independent, and  (\ref{uncnostra}) is minimal on the state(s) that are eigenvectors of $x_1$ with the highest (in absolute value) eigenvalue.
In particular, in Madore's FS it is \ $x_i\propto L_i$, \ $\bx^2\equiv 1$, \ and the spacial uncertainty (\ref{uncnostra}) 
coincides up to a factor with the aforementioned $(\Delta\bm{L})^2$; hence on the representation space $V_l$ it is minimized by the same  SCS, on which it amounts to
\be 
\label{uncMadore}
(\Delta \bm{x})^2_{\emph{min}}=\frac{2}{n+1}=\frac{1}{l+1}.
\ee
Using the results of \cite{newpaper} here we are going to show that on our fuzzy spheres $S^d_\Lambda$
\be\label{uncnostra'}
(\Delta \bm{x})^2_{\emph{min}}< \frac{C_d}{(\Lambda+1)^2},\quad\mbox{where }C_d=\begin{cases}
3.5\quad \mbox{if }d=1,\\
11\quad \mbox{if }d=2,
\end{cases}
\ee
and that the states minimizing $(\Delta \bm{x})^2$ make up a weak SCS. Its elements can be considered as the closest \cite{Perelomov} states  to pure classical states  - i.e. points - of $S^d$ , because they are in one-to-one correspondence with points of $S^d$, are optimally localized around the latter  and are mapped into each other by the symmetry group $O(D)$.
In the case $d=2$ the right-hand side goes to zero as $\Lambda\to\infty$ much faster than the uncertainty (\ref{uncMadore}) 
for all irreducible components appearing in the decomposition $\mathcal{H}_{\Lambda}=\bigoplus\nolimits_{l=0}^\Lambda V_l$, \  including the one 
$(\Delta \bm{x})^2_{\emph{min}}={1}/{(\Lambda+1)}$ \ corresponding to
the  highest $l$. In this sense the optimally localized states on our $S^2_\Lambda$ have a sharper  spacial  localization than  the CS on Madore FS\footnote{Of course a future,
more precise determination of  $(\Delta \bm{x})^2_{\emph{min}}$ will indicate an even sharper localization.}.
We are also going  to determine various strong SCS, in particular one with \ $(\Delta \bm{x})^2< 1/(\Lambda+1)$; \ the elements of the latter SCS are eigenvectors of a suitable component of the angular momentum, so that the corresponding states (rays or equivalently 1-dim projections) are in one-to-one correspondence with points of $S^d$, and the resolution of the identity holds also integrating just over the coset space $S^d$.

The plan of the paper is as follows. In section \ref{preliminari} we collect preliminaries: in section \ref{preliminari1} we recall some basic facts about the theory of Coherent States as treated in \cite{Perelomov} and its application to SO(3), leading in particular to (\ref{uncMadore}); in sections \ref{preliminari2}, \ref{preliminari2'} we respectively derive uncertainty relations (UR) on the commutative $S^1$, $S^2$ and briefly discuss coherent states on them; in section \ref{preliminari4} we explain how a 
tridiagonal Toeplitz matrix can be diagonalized. In sections \ref{CSD=2} and \ref{CSD=3} we respectively determine uncertainty relations, coherent and localized states on our fuzzy spheres $S^1_\Lambda$ and $S^2_\Lambda$: we first  recall the main features \cite{FiorePisacaneJGP18,FiorePisacanePOS18} of these $S^d_{\Lambda}$, then  (sections \ref{circleCS1} and \ref{sphereCS1}) we derive $O(D)$-invariant UR, introduce  classes of strong SCS  on them and approximately determine the corresponding $(\Delta \bm{x})^2,(\Delta \bm{L})^2$, finally we introduce and approximately determine the $O(D)$-invariant weak SCS  minimizing the spatial dispersion $(\Delta \bm{x})^2$ (sections \ref{circleCS2} and \ref{CS3}). 
Section \ref{Conclu} contains final remarks - including a detailed comparison of CS on
Madore FS and our $S^2_\Lambda$ -, outlook and
conclusions. In the Appendix (section \ref{Appendix}) we have concentrated some useful notions, lengthy computations and proofs.

\section{Further preliminaries}\label{preliminari}

\subsection{Basics about Coherent States}\label{preliminari1}

Coherent states (CS) were originally introduced in  quantum mechanics on $\RR^3$ as states \cite{Schroe26,Klauder,Glauber} saturating the   Heisenberg uncertainty relations (HUR) 
\ $\Delta x_i\Delta p_i\ge \hbar/2$ and mapped into each other by the Heisenberg-Weyl group; \
they make up an overcomplete set  
yielding a nice resolution of the identity. The latter properties are usually taken as minimal requirements 
\cite{Klauder-Skager85} for  defining CS in general:
a set of CS $\left\{\phi_l\right\}_{l\in\Omega}$ is a particular set of vectors of a Hilbert space $\mathcal{H}$, where $l$ is an element of an appropriate (topological) label space $\Omega$, such that the following properties hold:

\medskip
\begin{enumerate}
\item  \textbf{Continuity}: the vector $\phi_l $ is a strongly continuous function of the label $l$.
\item \textbf{Resolution of the identity}:  there exists on $\Omega$ an integration measure such that 
\be
I=\int_{\Omega} P_l \, dl, \qquad P_l:=
\phi_l  \langle \phi_l,\cdot\rangle \equiv|\phi_l\rangle\langle\phi_l|  ; \label{ResolId0}
\ee
\item or, at least, \ \textbf{Completeness}: \ \
$\overline{\mbox{Span}\left\{\phi_l:l\in\Omega\right\}}=\mathcal{H}$;
\end{enumerate}
the first two properties characterize a strong SCS, while the first and third a weak SCS.

A. M. Perelomov  and  R. Gilmore develop \cite{PerelomovCMP72,GilmoreAnPh72} the concept of CS when $\Omega$ is a Lie group $G$ acting  on a Hilbert space $\mathcal{H}$ via an unitary irreducible representation $T$ (see e.g. Perelomov's book \cite{Perelomov}). Actually, most arguments hold also if the group $G$ is not Lie. 
Fixed $\bm{\phi}_0\in\mathcal{H}$ Perelomov defines 
\ $\bm{\phi}_g:=T(g)\bm{\phi}_0 $ 
\ and the {\it coherent-state system}  $\{T, \bm{\phi}_0\}$ \ as
\be
\{T, \bm{\phi}_0\}:=\{ \bm{\phi}_g:=T(g)\bm{\phi}_0 \: |\: g\in G\}.               \label{PereloCS}
\ee
Clearly $\{T, \bm{\phi}_0\}=\{T, \bm{\phi}_g\}$ for all $g\in G$. \ The maximal subgroup $H$  of $G$ formed by elements $h$ fulfilling 
$$
\bm{\phi}_h 
=\exp{\left[i\alpha(h)\right]}\bm{\phi}_0,
$$
with some function $\alpha:H\rightarrow\mathbb{R}$, is called the isotropy subgroup for $\bm{\phi}_0$. Clearly, $g'=gh$ implies 
$$
\bm{\phi}_{g'} =T(g)T(h)\bm{\phi}_0=T(g)\exp{\left[i\alpha(h)\right]}\bm{\phi}_0=\exp{\left[i\alpha(h)\right]}\bm{\phi}_g,
$$
i.e. $\bm{\phi}_{g'}, \bm{\phi}_g$ belong to the same ray.
Therefore equivalence classes
$x(g):=\{g'=gh\: |\: h\in H\}$, i.e. elements  of the coset space $X:=G/H$, are in one-to-one correspondence with coherent rays, or equivalently with coherent 1-dimensional projections  (states):  hence we shall denote
$ P_g:=\bm{\phi}_g\langle\bm{\phi}_g,\cdot\rangle=P_{g'}$ also  as $P_x$.
A left-invariant measure $d\mu(g)$ on $G$ induces an invariant measure $dx$ on $X$.
$T$ is said square-integrable if $I_T\equiv\int_X|\langle\phi_0,T[g(x)]\phi_0\rangle|^2\,dx<\infty$ (this is automatically true if $G$, or at least  $X$, is compact, because then the volume of $X$ is finite); \  here $g(x)$ is any
(smooth) map from $X$ to $G$ such that $g(x)\in x$ [the result does not depend on the representative element in $x$ because it is invariant under the replacement
$g\mapsto gh$; $g(x)$ can be seen as a section of a $U(1)$-fiber bundle on $X$].
If $T$ is  square-integrable then the integral defining the operator
\ $B:=\int_XP_x\,dx$ \ is automatically convergent. 
From the identities $T(g')P_xT(g'{}^{-1})=P_{x'}$ (with $x':=g'x$)
and the invariance of $dx$ it follows that $T(g')B\,T(g'{}^{-1})=B$, and therefore
$B$ is central; then by Schur lemma there is $b\in \RR^+$ such that $B=bI$.
One can determine $b$ taking the mean value of both sides on $\phi_0$; one easily finds
\ $b\langle\phi_0,\phi_0\rangle=I_T$.
In general the set $\{\phi_{g(x)}\}_{x\in X}$ is overcomplete (this is certainly the case if $X$ is a {\it continuum}); one can extract  a basis out of it in many different ways.
Introducing the normalized integration measure $d\nu(x):=dx/b$ one finds
the first resolution of the identity in
\be
I=\int_XP_x\,d\nu(x),\qquad       I=\int_GP_g\,d\mu'(g);                                 \label{ResolId}
\ee
the second holds if $H$ has a finite volume $h$, and we define  $d\mu'(g):=d\mu(g)/bh$, so $\{T, \bm{\phi}_0\}$ is a strong SCS.  
In particular, Perelomov applies  (chpt. 4 in \cite{Perelomov}) these notions to the 
irreducible representation $(\pi_l,V_l)$ of $G=SU(2)$
selecting  a vector $\bm{\phi}_0$ that 
maximizes the  isotropy subalgebra $\b$
in the complex hull $sl(2,\CC)$ of the Lie algebra $su(2)$ 
and minimizes the
square dispersion $(\Delta\bm{L})^2$. As explained in the introduction, one
possible such $\bm{\phi}_0$ is the highest weight vector
$|l, l\rangle\in V_l$, i.e. the eigenvector of $L_3$ with the highest eigenvalue $l$ ($L_3|l, m\rangle= m|l, m\rangle$ with $|m|\le l$, in standard ket notation), which plays the role of vacuum (it is annihilated
by $L_+$) and has expectation values $\langle L_1\rangle=\langle L_2\rangle=0$, 
$(\Delta\bm{L})^2=(\Delta\bm{L})^2_{\emph{min}}=l$. 
Therefore these CS coincide with the socalled {\it coherent spin} \cite{Rad71} or {\it Bloch} states.
 By  the $SU(2)$ invariance of $(\Delta\bm{L})^2$,  all elements  \ $\bm{\phi}_g\in\{\pi_l, \bm{\phi}_0=|l, l\rangle \}$ \ - including $|l,-l\rangle\sim T(e^{i\pi L_1})|l,l\rangle$ - have  the same minimal dispersion and are eigenvectors of 
the ``annihilation operator" $L_+$. 
As the isotropy subgroup $H$ of $|l, l\rangle$
is that $SO(2)$ of rotations $e^{i\varphi L_3}$ around the $\vec{z}$-axis, the  states associated with this system are in one-to-one correspondence with the points of   $SO(3)/SO(2)=S^2$.
The latter sphere can be considered as the phase manifold for spin (angular momentum); these  coherent states are the closest to the classical ones on such a
sphere. 
Applying the rescaling (\ref{rescale}) we immediately find  that also in the Madore FS the space uncertainty is minimal on  the $|\bm{\phi}_g\rangle$'s and equal to (\ref{uncMadore}).

Out of the $\phi_g$'s only the vectors proportional to $|l,\pm l\rangle$ saturate (i.e. satisfy as equalities) 
 for all $i,j$ the uncertainty relations
$\Delta L_i\,\Delta L_j\ge |\varepsilon^{ijk}\langle L_k\rangle|/2$, which follow from the commutation relation $[L_i,L_j]=i\varepsilon^{ijk}  L_k $ (on them one has in addition
$\langle L_1\rangle=\langle L_2\rangle=0=\Delta L_3$, $|\langle L_3\rangle|=l$, $\Delta L_1=\Delta L_2=\sqrt{l/2}$). Incidentally, the authors in Ref. \cite{AraChaSal76}
consider also two alternative definitions of sets of optimally localized states: the set of ``intelligent states'', that saturate the uncertainty relation $\Delta L_1\,\Delta L_2\ge |\langle L_3\rangle|/2$, and the set of ``minimum uncertainty states'', for which $\Delta L_1\,\Delta L_2$ has a local minimum (note that then in general $\Delta L_1\,\Delta L_3$, $\Delta L_2\,\Delta L_3$ are not minimized). But neither one is invariant under arbitrary rotation, in contrast with the definition of Perelomov and of the present work; one can easily show (see e.g. 
\cite{Klauder}   pp. 27-28) that these states  are ``fewer''
than  the points of $S^2$, i.e cannot be put in one-to-one correspondence with the points of $S^2$, but just  of  a finite number of lines on $S^2$.

\subsection{Uncertainty relations and coherent states on commutative $S^1$}\label{preliminari2}

Let $x_1,x_2$ be Cartesian coordinates on $\RR^2$, $\partial_i\!\equiv\!\partial/\partial x_i$,   $L=-i(x_1\partial_2-x_2\partial_1)$ be the angular momentum
operator up to $\hbar$ . From  $[L,x_1]=ix_2$, $[L,x_2]=-ix_1$
one derives  in the standard way the  uncertainty relations (UR)
\bea
(\Delta L)^2(\Delta x_1)^2\ge \frac 14\langle x_2\rangle^2,\qquad 
(\Delta L)^2(\Delta x_2)^2\ge \frac 14\langle x_1\rangle^2,\qquad 
(\Delta L)^2(\Delta \bx)^2\ge \frac 14\langle \bx\rangle^2;         \label{HURS^1}
\eea
the third inequality is obtained summing the first two. These commutation relations and UR hold not only 
for the operators on $\Hi={\cal L}^2(\RR^2)$, but also for those on $\Hi={\cal L}^2(S^1)$. In the latter case
the $x_i$ fulfill the constraint $\bx^2\equiv x_1^2+x_2^2=1$, or equivalently $x_+x_-=1$, where $x_\pm:=x_1 \pm ix_2$,   whence $(x_+)^{-n}=(x_-)^n$, and the third inequality represents a lower bound for the  dispersion $\ \Delta L\, |\Delta \bx|$ in phase space; $L$ is the momentum along the circle. The inequalities  (\ref{HURS^1}) are therefore the  analog \cite{Lev76}
on the circle of the Heisenberg UR (we recall that adopting the azimuthal
angle $\varphi$ as the observable canonically conjugate  to $L$, $[\varphi,L]=i$,  would be inconsistent).
The  orthonormal basis $\B:=\{\psi_n\}_{n\in\ZZ}$ of ${\cal L}^2(S^1)$,
$\sqrt{2\pi}\psi_n:=e^{in\varphi}=  (x_+)^n$  consists of  eigenvectors of $L$, $L\psi_n=n\psi_n$,
while $x_\pm$ acta as ladder operators: $x_\pm\psi_n=\psi_{n\pm 1}$. \  These relations characterize the basic\footnote{The inequivalent unitary irreducible representation of $\A$ are parametrized by $\alpha\in[0,2\pi[$, entering  $L\psi_n=(n+\alpha)\psi_n$.} unitary irreducible representation $T$ of the $*$-algebra $\A$ of observables generated by $L,x_\pm$ fulfilling 
$[L,x_\pm]=\pm x_\pm, \:\: x_+x_-=x_-x_+=1, \:\: L^\dagger=L, \:\:  x_+^\dagger=x_-$.
The $\psi_n$ saturate the inequalities  (\ref{HURS^1}), because on them $(\Delta L)^2 = \langle x_1\rangle= \langle x_2\rangle=0$, while $(\Delta x_i)^2=1/2$; in appendix \ref{diverged=1} we show that in fact these are the only states saturating (\ref{HURS^1}).
The decomposition of the identity associated to $\B$ (first equality)
\be
I=\sum_{n} P_n=\int_{G/H}P_x\,d\mu(x), \qquad P_n=\psi_n\langle \psi_n,\cdot\rangle \label{ResolIdS^1}
\ee
thus involves all and only the states saturating (\ref{HURS^1}), i.e. is of the type
(\ref{ResolId0}) with labels $n\in\Omega\equiv \ZZ$; the second equality is explained once
we note that 
$\Hi={\cal L}^2(S^1)$ carries a unitary irreducible representation of the group 
\be
G:=\{(x_+)^n e^{i(a L+b)} \, |\, (a,b,n)\in\RR^2\times \ZZ\}\simeq U(1)\times U(1)\,\cross\ZZ
\label{G1S^1}
\ee
(consisting of  $*$-automorphisms  of the algebra of observables) with product rule
$$
(x_+)^ne^{i(aL+b)}\, (x_+)^{n'}e^{i(a'L+b')}=(x_+)^{n+n'}e^{i[(a+a')L+(b+b'+an')]};
$$ 
$e^{iaL}\psi(\varphi)=\psi(\varphi+a)$, i.e.
$e^{iaL}$ is the translation operator along the circle (it rotates $\varphi$ by an angle $a$), while $x_\pm\psi_m=\psi_{m\pm 1}$,
i.e. $x_\pm$ act as discretized boost operators in the (anti)clockwise direction.
$G$ acts transitively on the set of states saturating the HUR (\ref{HURS^1}), i.e. the eigenvectors of $L$. 
$H=\{e^{i(a L+b)}\}\simeq [U(1)]^2$ is the isotropy subgroup of $\psi_0$ (and of all other $\psi_n$),  and $G/H=\{(x_+)^n  \, |\, n\in  \ZZ\}$, hence integrating over $G/H$ amounts to summing over $n\in\ZZ$. In this broader sense $\{T, \psi_0\}$ is a strong SCS.

\subsection{Uncertainty relations and coherent states on commutative $S^2$}\label{preliminari2'}

From the commutation relation $[L_i,L_j]=i\varepsilon^{ijk}  L_k $  (for all $i,j$), valid   
on ${\cal L}^2(\RR^3)$  and ${\cal L}^2(S^2)$, 
one derives in the standard way the UR
\bea
\Delta L_1\,\Delta L_2\ge \frac 12 |\langle L_3\rangle|,\qquad
\Delta L_2\,\Delta L_3\ge \frac 12 |\langle L_1\rangle|,\qquad
\Delta L_3\,\Delta L_1\ge \frac 12 |\langle L_2\rangle|.\label{LUR3}
\eea
As already said, the set of coherent spin  states within  $\Hi=V_l$  is the subset of  states  minimizing  $(\Delta\bL)^2$.
Among them only $|l,l\rangle$,  $|l,-l\rangle$ saturate (\ref{LUR3}). Is there some UR which is saturated
by all coherent spin  states? We show in appendix \ref{proofpropoURS2} not  only that the answer is affirmative, 
but that such a UR is actually $l$-independent and valid on all of ${\cal L}^2(S^2)$:

\begin{teorema} The following uncertainty relation holds on 
${\cal L}^2(S^2)=\bigoplus_{l=0}^\infty V_l$
\bea
(\Delta\bL)^2\ge |\langle \bL\rangle| \qquad\Leftrightarrow \qquad
\langle \bL^2\rangle\ge 
|\langle \bL\rangle|\left(|\langle \bL\rangle|+1\right),  \label{LUR3'}
\eea
and is saturated by the spin coherent states 
 $\bm{\phi}_{l,g}=  \pi_l (g) |l,l\rangle\!\in\!  V_l\subset {\cal L}^2(S^2)$, $g\!\in\! SO(3)$,  $l\in\NN_0$. 
\label{propLUR3'}
\end{teorema}
\noindent
{\bf Remarks:}
\begin{enumerate}

\item As far as we know the theorem is new, albeit the proof is rather simple. 
One cannot obtain inequality (\ref{LUR3'}) directly from  (\ref{LUR3})
or the Robertson inqualities\footnote{Using (\ref{LUR3}) one can obtain the 
weaker inequality $(\Delta\bL)^2\ge |\langle \bL\rangle|\sqrt{3/4}$: \  (\ref{LUR3}) implies
the  inequalities $2\Delta L_1^2\,\Delta L_2^2\ge \langle L_3\rangle^2/2$,  $(\Delta L_1^4+\Delta L_2^4)/2\ge \langle L_3\rangle^2/4$ and the ones obtained   permuting $1,2,3$ cyclically; summing all of them  we obtain $(\Delta\bL)^4\ge \langle \bL\rangle^2|3/4$.}.

\item Summing  Perelomov's resolutions of the identities for all $V_l$  we obtain the
 resolution of the identity for  ${\cal L}^2(S^2)$
\be
I=\sum_{l=0}^\infty C_l\int_{SO(3)}\!\!\! \!\!\!  d\mu(g) \, P_{l,g},\qquad P_{l,g}=\bm{\phi}_{l,g}
\langle\bm{\phi}_{l,g},\cdot\rangle,\quad C_l=\frac{2l\!+\!1}{8\pi^2},\quad   \bm{\phi}_{l,g}:= T(g)Y^l_l;  \label{ResolIdS^2}
\ee
this holds also integrating over $S^2$ [instead of  $SO(3)$]  and replacing $C_l\mapsto 2\pi C_l$.
\end{enumerate}

From the commutation relation $[L_i,x_j]=i\varepsilon^{ijk}  x_k $ (for all $i,j$), valid   on ${\cal L}^2(\RR^3)$, and ${\cal L}^2(S^2)$, 
one derives in the standard way the UR
\bea
\Delta L_1\,\Delta x_2\ge \frac 12 |\langle x_3\rangle|,\qquad
\Delta L_1\,\Delta x_3\ge \frac 12 |\langle x_2\rangle|,\nn
\Delta L_2\,\Delta x_1\ge \frac 12 |\langle x_3\rangle|,\qquad
\Delta L_2\,\Delta x_3\ge \frac 12 |\langle x_1\rangle|,\label{HUR3}\\
\Delta L_3\,\Delta x_1\ge \frac 12 |\langle x_2\rangle|,\qquad
\Delta L_3\,\Delta x_2\ge \frac 12 |\langle x_1\rangle|.\nonumber
\eea
Relations (\ref{HUR3})  are  analogs of the Heisenberg UR (HUR), as the $L_i$ are the ``momentum'' components along the sphere. Alternative ones can be found e.g. in \cite{GoodmanGoh03}. 
We have not found in the literature works investigating whether they can be saturated. 

\subsection{Diagonalization of Toeplitz tridiagonal  matrices}\label{preliminari4}

A Toeplitz tri-diagonal matrix is a $n\times n$ matrix of the form
\be
P_n\left(a,b,c\right):=\left(
\begin{array}{cccccccc}
a&b&0&0&0&0&0&0\\
c&a&b&0&0&0&0&0\\
0&c&a&b&0&0&0&0\\
\vdots&\vdots&\vdots&\vdots&\ddots&\vdots&\vdots&\vdots\\
0&0&0&0&\cdots&a&b&0\\
0&0&0&0&\cdots&c&a&b\\
0&0&0&0&\cdots&0&c&a\\
\end{array}
\right);                     \label{Toeplitz}
\ee
its eigenvalues are (see e.g. \cite{Noschese} p. 2-3)
$$
\lambda_h=a+2\sqrt{bc}\cos{\left(\frac{h\pi}{n+1}\right)},\quad h=1,\cdots,n
$$
and  the corresponding eigenvectors $\chi^{h}$ are columns with the following components
$$
\chi^{h,k}=\left(\frac{c}{b}\right)^{\frac{k}{2}}\sin{\left(\frac{hk\pi}{n+1}\right)},\qquad h,k=1,2,\cdots,n,
$$
up to normalization. 
In the symmetric case ($b=c$) all eigenvalues are real and  the highest one is clearly 
$\lambda_1$; the norm of $\chi^1$ is  easily computed:
$$
\left\Vert\chi^1\right\Vert^2=\sum_{k=1}^{n}\sin^2{\left(\frac{k\pi}{n+1}\right)}=\frac{n+1}{2}. 
$$

\section{Coherent and localized states on the fuzzy circle $S^1_\Lambda$}
         \label{CSD=2}

We first recall how  $S^1_\Lambda$ is defined. In a suitable orthonormal basis $\B:=\{\psi_\Lambda,\psi_{\Lambda-1},...,\psi_{-\Lambda}\}$ of the Hilbert space $\Hi_\Lambda$ consisting of eigenvectors of the angular momentum $L\equiv L_{12}$,
\be
L \psi_n=n\psi_n,
\ee
the action of the noncommutative coordinates \ $x_\pm:=x_1 \pm ix_2$
 \ of the fuzzy circle $S^1_\Lambda$ read\footnote{We have changed conventions with respect to  \cite{FiorePisacaneJGP18}:
the  $x_i$ ($i=1,2$) as defined here equal the $\xi^i=\overline{x}^i/a$ of \cite{FiorePisacaneJGP18}
where \ $a=1\!+\!\frac 94\frac{1}{\sqrt{2k}}\!+\!O\!\left(\!\frac{1}{k}\!\right)$ is just a normalization  factor; the $x_\pm$ as defined here equal  $\sqrt{2}\xi^\pm=\sqrt{2}\overline{x}^\pm/a$ of \cite{FiorePisacaneJGP18}.}
\bea
x_+\psi_n=
 b_{n+1}\psi_{n+1} ,  \qquad x_-\psi_n=b_n\psi_{n-1} ,  \qquad 
b_n:=\left\{\!\!\ba{ll}\displaystyle    \sqrt{1\!+\!\frac{n(n \!-\! 1)}{k}} \:\: &
\mbox{if }1\!-\!\Lambda\leq  n\leq\Lambda, \\[10pt]
0 & \mbox{otherwise}.
\ea\right. 
      \label{defLxiD=2}
\eea
Note that \ 
$b_{-\Lambda}=b_{\Lambda+1}=0$, \ \ $b_n=b_{1-n}$ \ if $\Lambda+1\ge n\ge 0$.

$L ,x_+,x_-$ and $\bx^2:=x_1 ^2+x_2^2=(x_+x_-+x_-x_+)/2$ fulfill the $O(2)$-equivariant  relations
\be
\left[L , x_{\pm}\right]=\pm x_\pm,\quad x_+{}^\dagger=x_-, \qquad L ^\dagger=L,\label{commrelD=2'}
\ee
\be
\left[x_+,x_-\right]=-\frac{2L }k+\left[1\!+\!\frac {\Lambda(\Lambda\!+\!1)}k\right]\!\left(\widetilde P_{\Lambda}\!-\!\widetilde P_{-\Lambda}\right),\label{y+y-}
\ee
\be
\bx^2=  1+\frac{L ^2}{k} -
\left[1\!+\!\frac {\Lambda(\Lambda\!+\!1)}{k}\right]
\frac{\widetilde P_{\Lambda}\!+\!\widetilde P_{-\Lambda}}2,          \label{defR2D=2}
\ee
\be
\prod\limits_{m=-\Lambda}^{\Lambda}\!\!\left(L \!-\!mI\right)=0, \qquad \left(x_\pm\right)^{2\Lambda+1}=0 .\label{commrelD=2}
\ee
Here $\widetilde{P}_m$ is the projection over the $1$-dim subspace spanned by $\psi_m$, and  $k$ is a function of $\Lambda$ fulfilling (\ref{kineq}).
We point out that: 

\begin{itemize}

\item $\bx^2\neq 1$, but it is a function of $L^2$, hence the $\psi_m$ are
its eigenvectors; its eigenvalues  
(except on $\psi_{\pm\Lambda}$)
are close to 1, slightly grow with $|m|$ and collapse to 1 as $\Lambda\to \infty$.

\item The ordered monomials  $x_+^h\,L ^l\, x_-^n$ [with degrees $h,l,n$
bounded by (\ref{commrelD=2'})-\ref{commrelD=2}] make up a 
 basis of the $(2\Lambda\!+\!1)^2$-dim  vector space 
underlying the algebra of observables  \ $\A_\Lambda\!:=\!End(\Hi_\Lambda\!)$
\ (the $\widetilde{P}_m$ themselves can be expressed as polynomials in $L $). 

\item $x_+ ,x_-$ generate the $*$-algebra $\A_\Lambda$,
because also $L $ can be expressed as a non-ordered polynomial in $x_+,x_-$. 

\item Actually  there are $*$-algebra isomorphisms $\A_{\Lambda}$ 
 \be
\A_{\Lambda}\simeq M_N(\CC)\simeq\pi_\Lambda[Uso(3)], \qquad 
N=2\Lambda\!+\!1,                                 \label{isomD2}
\ee
where $\pi_{\Lambda}$ is the $N$-dimensional unitary irreducible representation  of $Uso(3)$.
The latter is characterized by the condition $\pi_{\Lambda}(C)=\Lambda(\Lambda+1)$, where  $C=E_aE_{-a}$ is the Casimir (sum over $a\in\{+,0,-\}$), and $E_a$
make up the Cartan-Weyl basis of $so(3)$,
\be
[E_+,E_-]=E_0,\qquad [E_0,E_\pm]=\pm E_\pm,\qquad E_a^\dagger=E_{-a}.
 \label{su2rel}
\ee
In fact we can realize \  $L ,x_+, x_-$ \ by setting \cite{FiorePisacaneJGP18}
(we simplify the notation dropping $\pi_{{{\Lambda}}}$)
\be
\ba{c}
L=E_0, \qquad  x_\pm=f_{\pm}(E_0)E_\pm,\\[10pt]
\displaystyle  f_{+}(s)=\sqrt{\frac{1\!+\!s(s\!-\!1)/k}{{{\Lambda}}({{\Lambda}}+1)\!-\!s(s\!-\!1)}}= f_{-}(s-1),
\ea\label{transfD2}\
\ee
i.e. in a sense the $x_\pm$ are $E_\pm$ (which play the role of $x_\pm$ in  Madore FS) squeezed in the $E_0$ direction; one can easily check (\ref{commrelD=2'}-\ref{commrelD=2})  using (\ref{azioneL}), 
with $L_a,l,m$ resp.  replaced   by $E_a,\Lambda,n$.
Hence \ $\pi_{\Lambda}(E_+),\pi_{\Lambda}(E_-)$ \  are generators of 
$\A_\Lambda$ alternative to \ $x_+,x_-$.

\item
The group $Y_\Lambda\simeq SU(2\Lambda\!+\!1)$ of $*$-automorphisms  of $\A_\Lambda$ is inner and includes a subgroup $SO(3)$ {\it independent of $\Lambda$}
(acting irreducibly via $\pi_\Lambda$) and a subgroup $O(2)\subset SO(3)$  corresponding to orthogonal transformations (in particular, rotations) of the coordinates $x_i$, which play the role of isometries of $S^1_\Lambda$.

\item In the limit $\Lambda\to \infty$ \ 
dim$(\Hi_{\Lambda})\!\to\!\infty$,  \ and we recover quantum mechanics (QM) on the circle $S^1$ as sketched in section \ref{preliminari2} 
 (see \cite{FiorePisacaneJGP18} for details).

\end{itemize}
As in the commutative case we define \ $\left\langle\bx\,\right\rangle^2\!:=\!\langle x_1\rangle^2\!+\!\langle x_2\rangle^2$ and find 
$\left\langle\bx\,\right\rangle^2\!=\!\langle x_+\rangle\langle x_-\rangle\!=\!|\langle x_+\rangle|^2$.

\subsection{$O(2)$-invariant UR and strong SCS  on $S^1_\Lambda$}
\label{circleCS1}

We first note that, since relations (\ref{commrelD=2'}) are as in the commutative case,
 the ``Heisenberg'' UR (\ref{HURS^1}) hold,  the eigenvectors 
$\psi_n$ of $L$ make up again a set of states saturating (\ref{HURS^1}), 
because on them $\left(\Delta L\right)^2= \langle x_1\rangle= \langle x_2\rangle=0$, while 
$$
(\Delta x_i)^2=\left\{\!\!\ba{l}\frac 12\left(1+\frac{n^2}{k}\right) ,\\[6pt]
\frac 14\left[1+\frac{\Lambda(\Lambda\!-\!1)}{k}\right],
\ea\right.\qquad 
(\Delta \bx)^2=\left\{\!\!\ba{ll}1+\frac{n^2}{k} \qquad & \mbox{if } |n|<\Lambda,\\[6pt]
\frac 12\left[1+\frac{\Lambda(\Lambda\!-\!1)}{k}\right] \qquad & \mbox{if } |n|=\Lambda.
\ea\right.
$$
The first resolution of the identity in (\ref{ResolIdS^1})  still holds, 
\be
I=\sum_{n} P_n=\int_{G/H}P_x\,d\mu(x), \qquad P_n=\psi_n\langle \psi_n,\cdot\rangle,                                        \ee
provided
$n$ runs over $\Omega\equiv\{-\Lambda,1\!-\!\Lambda,..., \Lambda\}$ instead of $\ZZ$. For the second one  
 to be valid one should replace  $\ZZ$ by $\ZZ_{2\Lambda+1}$   in the definition (\ref{G1S^1}) of $G$, more precisely
replace $(x_+)^n$ by $u^n$, where the unitary operator $u$ is defined by 
$u\psi_{\Lambda}=\psi_{-\Lambda}$, $u\psi_n=\psi_{n+1}$  otherwise.
Such a $G$  is a subgroup of the group of $*$-automorphisms of $\A_{\Lambda}$.
In appendix \ref{diverged=1} we show that   in $\Hi_\Lambda$ again
only  the $\psi_n$ saturate all of the  inequalities of (\ref{HURS^1}). Nevertheless,  there is a whole family (parametrized by $\mu\in\RR$) of complete sets of states saturating (\ref{HURS^1})$_1$ alone. 
These states are eigenvectors of $a_1^\mu:=L-i\mu x_1$ (we explicitly determine them for $\Lambda=1$), and the family interpolates between the set of eigenvectors of $L$ and the set of eigenvectors of $x_1$.

In the commutative case the spacial 
uncertainties $\Delta x_1, \Delta x_2$ can be simultaneously as small as we wish.
In the fuzzy case even the Robertson UR
$$
4\left(\Delta x_1\right)^2 \left(\Delta x_2\right)^2\ge  \left\langle L'\right\rangle^2+\langle x_1x_2+x_2x_1\rangle^2,
\qquad L':=-\frac{L }k+\left[1\!+\!\frac {\Lambda(\Lambda\!+\!1)}k\right]\!\frac{\widetilde P_{\Lambda}\!-\!\widetilde P_{-\Lambda}}2, 
$$
which follows from (\ref{y+y-}) and is slightly stronger than the Schr\"odinger UR,
is not particularly stringent, in that the right-hand side vanishes on a large class of states\footnote{In fact, on the generic vector
$\bm{\chi}=\sum_{m=-\Lambda}^{\Lambda}{\chi_m\psi_m}$ one finds
$\left\langle L'\right\rangle_{\bm{\chi}}=\sum_{m=1}^{\Lambda-1}[|\chi_{-m}|^2\!-\!|\chi_{m}|^2]m/k
+[|\chi_{\Lambda}|^2\!-\!|\chi_{-\Lambda}|^2][1/2\!+\!\Lambda(\Lambda \!-\!1)/2k]$, which vanishes e.g. if 
$|\chi_{-m}|\!=\!|\chi_{m}|$ for all $m$, and
$\langle x_1x_2+x_2x_1\rangle=\langle x_+^2-x_-^2\rangle/2i$. 
which vanishes if e.g. all $\chi_m\in\RR$, so that $\langle x_+^2\rangle$
is real.}, hence does not exclude that either $\Delta x_1$ or  $\Delta x_2$ vanish.
However,  we will see that the latter cannot vanish simultaneously,  
because $(\Delta\bx)^2$ is bounded from below (see section \ref {circleCS2}).

We now apply  (\ref{PereloCS}) adopting \ $T\!=\!\pi_\Lambda$ \ and as a  $G$ not 
$SO(3)$ (the largest $\Lambda$-independent subgroup of the group of $*$-automorphism of $\A_\Lambda$), but its subgroup $G=SO(2)$; hence $\Hi_\Lambda$ carries a {\it reducible} representation of $G$, so that completeness and resolution of the identity are not automatic.
Consider a generic unit vector  $\bomega=\sum_{m=-\Lambda}^{\Lambda}{\omega_m\psi_m}$ and let 
$$
\bomega_\alpha:=e^{i\alpha L}\bomega=\sum_{m=-\Lambda}^{\Lambda}
e^{i\alpha m}\omega_m\psi_m,\qquad\qquad P_\alpha:=\bomega_\alpha\langle\bomega_\alpha,\cdot\rangle,
$$
($\bomega_0\equiv \bomega$). The system 
$A:=\{\bomega_\alpha\}_{\alpha\in[0,2\pi[}$ is complete
provided $\omega_m\neq 0$ for all $m$ (then it is also overcomplete). Defining
$B:=\int_0^{2\pi}\!\! d\alpha \,P_\alpha$ one finds
$$
B\psi_n=\overline{\omega_n}\int_0^{2\pi}\!\!\!\!\! \bomega_\alpha e^{-i\alpha n}\,d\alpha
=\overline{\omega_n}\sum_{m=-\Lambda}^{\Lambda}\omega_m\psi_m\int_0^{2\pi}\!\!\!\!\!  e^{i\alpha (m-n)}\,d\alpha=2\pi|\omega_n|^2\psi_n,
$$
implying $B=\sum_{n=-\Lambda}^{\Lambda}2\pi|\omega_n|^2\widetilde P_n$; this is proportional
to the identity only if $|\omega_n|^2$ is independent of $n$ and therefore (since 
$\bomega$ is normalized) if $|\omega_n|^2=1/(2\Lambda\!+\!1)$. Setting
$\omega_n=e^{i\beta_n}/\sqrt{2\Lambda\!+\!1}$ we find the following resolutions of the identity, parametrized by \ $\beta\in(\RR/2\pi\ZZ)^{2\Lambda\!+\!1}$:
\be
I=\frac{2\Lambda\!+\!1}{2\pi}\int_0^{2\pi}\!\! d\alpha \,P^\beta_\alpha,
\qquad\qquad P_\alpha^\beta:=\bomega_\alpha^\beta\langle\bomega_\alpha^\beta,\cdot\rangle,\qquad \bomega_\alpha^\beta:=\sum_{m=-\Lambda}^{\Lambda}
\frac{e^{i(\alpha m+\beta_m)}}{\sqrt{2\Lambda\!+\!1}}\psi_m.      \label{IdResol} 
\ee
By choosing $\beta_{-m}=\beta_m$ the strong SCS $\{\bomega_\alpha^\beta\}$ is fully $O(2)$-equivariant, 
because is mapped 
into itself also by the unitary transformation $\psi_m\mapsto \psi_{-m}$ that corresponds to 
the transformation of the coordinates (with determinant -1) \  $(x_1,x_2)\mapsto(x_1,-x_2)$.
We now look for the $\beta$ minimizing $(\Delta \bx)^2$.  \ In appendix \ref{Lulu}
we show that on the states $\bomega_\alpha^\beta$
\bea
\langle L\rangle=0, \qquad\qquad \left(\Delta L\right)^2=\langle L^2\rangle=\frac {\Lambda(\Lambda\!+\!1)}{3}
\quad\mbox{for all }\alpha,\beta,\label{Lphiab}\\
\langle\bx^2\rangle 
\leq\frac {2\Lambda }{2\Lambda\!+\!1}+ \frac {2(\Lambda\!-\!1)\Lambda(\Lambda\!+\!1)}{3(2\Lambda\!+\!1)k},\qquad
\langle x_+\rangle=\frac {e^{- i\alpha}}{2\Lambda\!+\!1} \!\!\sum\limits_{m=1-\Lambda}^{\Lambda}\!\!\!e^{i(\beta_{m-1}-\beta_m)}b_m.      
  \label{utile2'}
 \eea
Therefore $\left\langle\bx\,\right\rangle^2=|\langle x_+\rangle|^2$ is maximal, and $\left(\Delta{\bx}\right) ^2=\langle\bx^2\rangle -\left\langle\bx\,\right\rangle^2$ is minimal, if $\beta=0$; then 
\bea
\langle x_+\rangle_{\bphi_\alpha} 
=  \frac {2\,e^{- i\alpha}}{2\Lambda\!+\!1}   \sum\limits_{m=1}^{\Lambda} b_m,  
\qquad\qquad \left(\Delta\bx\right) ^2
<\frac 1{\Lambda+1}
\left(\frac 12+\frac{1}{3\Lambda}\right)\overset{\Lambda\ge 2}\le  \frac 2{3(\Lambda+1)}     \label{utileb} 
 \eea
where $\bphi_\alpha:=\bomega_\alpha^0$;  in particular $\langle x_2\rangle_{\bphi}=0$,  
$\langle x_1\rangle_{\bphi}=\langle x_+\rangle_{\bphi}\in\RR$,
where $\bphi:=\bphi_0=\bomega_0^0$. We shall denote as
${\cal S}^1:=\{\bphi_\alpha\}_{\alpha\in[0,2\pi[}$ the corresponding strong SCS.

The $\bomega_\alpha^\beta$ have no limit in ${\cal L}^2(S^1)$ as  $\Lambda\to\infty$,
since all their components in the canonical basis $\{\psi_n\}_{n\in\ZZ}$ go to zero;
the renormalized $\sqrt{2\Lambda\!+\!1}\bphi_\alpha/2\pi$ have at least a limit in 
the space of distributions, more precisely go to $\delta_\alpha$,
where $\delta_\alpha$ is the Dirac $\delta$ on the circle centered at angle $\varphi=\alpha$.

\subsection{$O(2)$-invariant weaks SCS  on $S^1_\Lambda$ minimizing $(\Delta \bm{x})^2$}
\label{circleCS2}

As $(\Delta \bm{x})^2$ is $O(2)$-invariant, so is the set ${\cal W}^1$ of states on $S^1_\Lambda$ minimizing 
$(\Delta \bm{x})^2$. Therefore one can first look for a state 
$\underline{\bm{\chi}}\in{\cal W}^1$
such that $\langle x_2\rangle=0$, and then recover the whole ${\cal W}^1$ as
${\cal W}^1=\{\underline{\bm{\chi}}_\alpha:=e^{i\alpha L}\widehat{\bm{\chi}} \, |\, \alpha\in[0,2\pi[\}$.
This is an $O(2)$-invariant, overcomplete set of states (i.e. a weak SCS) in one-to-one correspondence with the points of the circle. The  determination  in closed form  of $\underline{\bm{\chi}},{\cal W}^1$ for general $\Lambda$ is presumably not possible.
Since it is $\bx^2=1+O(1/\Lambda^2)$ (except on $\bpsi_{\pm \Lambda}$), we expect that the eigenstate $\widehat{\bm{\chi}}$
of $x_1$ with highest eigenvalue (or the eigenstate  with  opposite eigenvalue) 
approximates $\underline{\bm{\chi}}$ at order $O(1/\Lambda^2)$. But also the
 determination  in closed form of such an eigenvector  is presumably not possible.
Here we content ourselves with giving $\underline{\bm{\chi}},\widehat{\bm{\chi}}$ for $\Lambda=1$ and finding for general $\Lambda$ a  set of states having a
smaller $(\Delta \bm{x})^2$ than that of the $\bphi_\alpha$
of the previous subsection, more precisely going to zero as $1/\Lambda^2$; this is done
with the help of the results  of \cite{newpaper}, where  a detailed study of the $x_i$-eigenvalue problem is carried out. 
 
When $\Lambda=1$ normalized eigenvectors and eigenvalues of $x_1$ are given by
\bea
\bm{\chi}_0=\frac {\psi_{-1}\!-\!\psi_1}{\sqrt{2}},\quad x_1\bm{\chi}_0=0,
\qquad \bm{\chi}_\pm=\frac {\psi_{-1}\!\pm\!\sqrt{2}\psi_0\!+\!\psi_1}{2},\quad x_1\bm{\chi}_\pm=\pm\frac {\sqrt{2}}2\bm{\chi}_\pm.
\eea
One easily checks that on $\widehat{\bm{\chi}}\equiv\bm{\chi}_+$  it is 
$\langle\bx^2\rangle=3/4$, $\langle x_+\rangle=\sqrt{2}/2$, and therefore
$(\Delta \bx)^2=1/4$. On the other hand in section \ref{Lulu} we show that 
$(\Delta \bx)^2$ is slightly smaller on $\underline{\bm{\chi}}$:
\bea
\underline{\bm{\chi}}=\frac{\sqrt{5}}4\left[\psi_{-1}\!+\!\psi_1\right]+\frac{\sqrt{3}}{\sqrt{8}}\psi_0\qquad\Rightarrow\qquad (\Delta \bx)^2=(\Delta\bx)^2_{min}=\frac 7{32}.        \label{underlinechi}
\eea

For general $\Lambda$, on the basis $\B_\Lambda$ of $\Hi_\Lambda$  the operator 
$x_1$ is represented by the $(2\Lambda\!+\!1)\times(2\Lambda\!+\!1)$ matrix
$$
X^\Lambda=\frac 12\left(\!
\begin{array}{cccccccc}
0&\!b_{\Lambda}\! &0&0&0&0&0&0\\
\!b_{\Lambda}\!  &0&\!b_{\Lambda-1}\! &0&0&0&0&0\\
0&\! b_{\Lambda-1}\! & 0&\!b_{\Lambda-2}\! &0&0&0&0\\
\vdots&\vdots&\vdots&\vdots&\ddots&\vdots&\vdots&\vdots\\
0&0&0&0&\cdots&\!b_{2-\Lambda}\! &0&\!b_{1-\Lambda}\! \\
0&0&0&0&\cdots&0&\!b_{1-\Lambda}\! &0 
\end{array}
\!\right)         =X^\Lambda_0+O\!\left(\!\frac {1}{\Lambda^2}\!\right)\!,\qquad X_0^\Lambda:=\frac 12 P_{2\Lambda+1}(0,1,1)
$$
 [see (\ref{Toeplitz})]. The spectrum $\Sigma^\Lambda_0$ of $X^\Lambda_0$ is $\{\cos[\pi n/(2\Lambda\!+\!2)]\}_{n=1,2,...,2\Lambda+1}$ (see section \ref{preliminari4});  $\Sigma^{\Lambda+1}_0$, $\Sigma^\Lambda_0$ interlace,
i.e. between any two subsequent eigenvalues in  $\Sigma^{\Lambda+1}_0$ there is exactly one in $\Sigma^\Lambda_0$, and $\Sigma^\Lambda_0$ becomes uniformly dense in $[-1,1]$ as $\Lambda\to \infty$. 
In \cite{newpaper} we  show that the same properties hold true also for   $X^\Lambda\simeq x_1$, by
studying its spectrum.
Here as a first good estimate of  $\widehat{\bm{\chi}}$ we take the eigenvector $\bm{\chi}$ of the Toeplitz matrix $X^\Lambda_0$ with the maximal eigenvalue $\lambda_M=\cos\left[\pi/(2\Lambda\!+\!2)\right]$. The associated  
 $(\Delta\bx)^2_{\bm{\chi}}$, which is a first good estimate of  $(\Delta\bx)^2_{min}$ and goes to zero as $1/\Lambda^2$, fulfills  (see appendix \ref{Lulu}) 
\bea
\left(\Delta{\bx}\right)^2_{\bm{\chi}}< \frac{3.5}{(\Lambda+1)^2}.            \label{Deltax2qminS^1_L}
\eea

\section{Coherent and localized states on the fuzzy sphere $S^2_\Lambda$}
  \label{CSD=3}


We first recall how  $S^2_\Lambda$ is defined. 
We use two related sets of  angular momentum and space coordinate operators: the hermitean ones
$\left\{L_i\right\}_{i=1}^3$ and $\left\{x_i\right\}_{i=1}^3$,  and the hermitean conjugate ones
$\left\{L_a\right\}$, $\left\{x_a\right\}$ (here $a=0,+,-$), which are obtained from the former as follows\footnote{We have changed conventions with respect to  \cite{FiorePisacaneJGP18}:
the  $x_i,L_i$ ($i=1,2,3$) as defined here respectively equal the $\overline{x}^i,\overline{L}_i$ of \cite{FiorePisacaneJGP18}; 
the $x_\pm,L_\pm$ as defined here respectively equal  $\sqrt{2}\overline{x}^\pm$, $\sqrt{2}\overline{L}_\pm$
 of \cite{FiorePisacaneJGP18}.}:
$$
L_{\pm}:=L_1\pm iL_2,\quad\quad L_0:=L_3,\quad\quad x_{\pm}:=x_1\pm ix_2,\quad\quad x_0:=x_3.
$$
The square distance from the origin can be expressed as $\bx^2:=x_{i}x_{i}=x_0^2+(x_+x_-+x_-x_+)/2$.
As a preferred orthonormal basis $\B_\Lambda$ of the carrier Hilbert space $\Hi_\Lambda$ we adopt one 
consisting of eigenvectors of \ $L_3$, \  $\bL^2=L_{i}L_{i}=L_0^2+(L_+L_-+L_-L_+)/2$,
\be
\B_\Lambda:=\left\{\bm{\psi}_l^m\right\}_{l=0,1,...,\Lambda; \ m=-l,...,l},\qquad
\bL^2\bm{\psi}_l^m=l(l+1)\bm{\psi}_l^m,\qquad L_3\bm{\psi}_l^m=m\bm{\psi}_l^m.
\ee
On the $\bm{\psi}_l^m$ the $L_a,x_a$ act as follows:
\be\label{azioneL}
L_0\bm{\psi}_l^m=m\,\bm{\psi}_l^m,\quad
L_{\pm}\bm{\psi}_l^m=\sqrt{(l\!\mp\! m)(l\!\pm\! m\!+\!1)}\bm{\psi}_l^{m\pm 1}\!,
\ee
\bea\label{azionex}
x_a\bm{\psi}_{l}^m=\left\{\!\!
\ba{ll}
c_l A_{l}^{a,m}\bm{\psi}_{l-1}^{m+a}+
c_{l+1} B_{l}^{a,m} \bm{\psi}_{l+1}^{m+a
}&\mbox{ if }l<\Lambda,\\[8pt]
c_lA_l^{a,m}\bm{\psi}_{\Lambda-1}^{m+a}&\mbox{ if }
l=\Lambda,\\[8pt]
0&\mbox{otherwise,}
\ea
\right. 
\eea
where 
\bea 
\label{Clebsch}
A_l^{0,m}=\sqrt{\frac{(l+m)(l-m)}{(2l+1)(2l-1)}}\hspace{0.3cm},\hspace{0.3cm} A_l^{\pm,m}=\pm\sqrt{\frac{(l\mp m)(l\mp m-1)}{(2l-1)(2l+1)}}\hspace{0.3cm},\hspace{0.3cm} B_l^{a,m}=A_{l+1}^{-a,m+a},\\
c_l:= \sqrt{1+\frac{l^2}{k}}\qquad 1\le l\le \Lambda,\qquad c_0=c_{\Lambda+1}=0.           \label{defcl}
\eea
where $k$ is a function of $\Lambda$ fulfilling (\ref{kineq}).
The $L_i,x_i$  fulfill the following $O(3)$-equivariant relations: 
\bea
&& x_i^{\dag}=x_i, \qquad 
L_i ^{\dag}=L_i, \qquad [L_i,x_j]=i\varepsilon^{ijh}x_h, \qquad 
\left[\,L_i,L_j\right]=i\varepsilon^{ijh}L_h,\qquad x_iL_i=0, \label{rf3D4}\\[8pt]
&& [x_i,x_j]=i\varepsilon^{ijh}\left(\!-\frac{I}{k}+K\widetilde{P}_{\Lambda}\!\right)L_h, \hspace{1cm}
\bx^2= 1
+\frac{\bL^2\!+\!1}{k}-\left[1+\frac{(\Lambda\!+\!1)^2}{k}\right]\frac{\Lambda\!+\!1}{2\Lambda\!+\!1}\widetilde P_{\Lambda},   \qquad                           \label{xx}\\[4pt]
&& \prod_{l=0}^{\Lambda}\left[\bL^2-l(l+1)I\right] =0,\qquad
\prod_{m=-l}^{l}{\left(L_3-mI\right)}\widetilde{P}_l=0,\qquad \left(x_{\pm}\right)^{2\Lambda+1}=0; \label{rf3D3}
\eea
here $K=\frac{1}{k}+\frac{1+\frac{\Lambda^2}{k}}{2\Lambda+1}$, $\widetilde{P}_l$ is the projection on the  $\bL^2=l(l+1)$ eigenspace.
We point out that: 
\begin{itemize} 

\item  $\bx^2\neq 1$; but it is a function of $\bL^2$, hence the $\bpsi_l^m$ are
its eigenvectors; and,  for each fixed $\Lambda$, its eigenvalues  (except when $l=\Lambda$) are close to 1, slightly grow with $l$ and collapse to 1 as $\Lambda\to \infty$.

\item  The ordered monomials in $x_i,L_i$ [with degrees  bounded by (\ref{rf3D4}-\ref{rf3D3})] make up a 
 basis of the $(\Lambda\!+\!1)^4$-dim  vector space \ $\mathcal{A}_{\Lambda}\!:=\!End(\Hi_\Lambda\!)\!\simeq\! M_{(\Lambda+1)^2}(\CC)$, 
because the $\widetilde{P}_l$ themselves can be expressed as polynomials in $\bL^2$.

\item The $x_i$ {\it generate} the $*$-algebra $\mathcal{A}_{\Lambda}$, because
also the $L_i$ can be expressed as non-ordered polynomials in the $x_i$.

\item Actually there are $*$-algebra isomorphisms 
 \be
\A_{\Lambda}\simeq M_N(\CC)\simeq\bpi_\Lambda[Uso(4)], \quad N:=(\Lambda\!+\!1)^2.             \label{isomD4}
\ee
where $\bpi_{\Lambda}$ is the unitary vector representation 
of $Uso(4)
$   on a Hilbert space 
${\bf V}_{\Lambda}
$, which is
characterized by the conditions $\bpi_{\Lambda}(C)=\Lambda(\Lambda+2)$, 
$\bpi_{\Lambda}(C')=0$ on the quadratic Casimirs.
In terms of the Cartan-Weyl basis  $\{L_{\lambda \mu }\}$ ($\lambda,\mu\in\{1,2,3,4\}$)  of $so(4)$,
\be
[L_{\lambda \mu },L_{\nu \rho }]=i[\delta_{\lambda \nu }L_{\mu \rho }-\delta_{\lambda \rho }L_{\mu \nu }-\delta_{\mu \nu }L_{\lambda \rho }+\delta_{\mu \rho }L_{\lambda \nu }], \qquad L_{\lambda \mu }^\dagger=L_{\lambda \mu }=-L_{\mu \lambda },
 \label{so4rel}
\ee
$C=L_{\mu \nu }L_{\mu \nu }$,  $C'=\varepsilon^{\lambda \mu \nu \rho }L_{\lambda \mu }L_{\nu \rho }$  (sum over repeated indices).
To simplify the notation we drop $\bpi_{\Lambda}$. 
In fact one can realize \  $L_i\, ,x_i$, \ $i\in\{1,2,3\}$, by  \ setting \cite{FiorePisacaneJGP18}
\be
\ba{rcl}
 L_i&=& \displaystyle\frac 1{2i}\varepsilon^{ijk}L_{jk}, \qquad \qquad  x_i=
g^*(\lambda)\, L_{4i}\,g(\lambda),\\[10pt]
 g(l) &=& \displaystyle\sqrt{
\frac{\Gamma\!\left(\frac {\Lambda\!+\!l}2\!+\!1\right)\Gamma\!\left(\frac {\Lambda\!-\!l\!+\!1}2\right)}
{\Gamma\!\left(\frac {\Lambda\!+\!1\!+\!l}2\!+\!1\right)\Gamma\!\left(\frac {\Lambda\!-\!l}2\!+\!1\right)}
\frac{\Gamma\!\left(\frac l2\!+\!1\!+\!\frac{i\sqrt{k}}2\right)\Gamma\!\left(\frac l2\!+\!1\!-\!\frac{i\sqrt{k}}2\right)}
{\sqrt{k}\:\Gamma\!\left(\frac {l\!+\!1}2\!+\!\frac{i\sqrt{k}}2\right)\Gamma\!\left(\frac {l\!+\!1}2\!-\!\frac{i\sqrt{k}}2\right)}}\\[22pt]
&=&  \displaystyle\sqrt{\frac{\prod_{h=0}^{l-1}(\Lambda\!+\!l\!-\!2h)}{\prod_{h=0}^l(\Lambda\!+\!l\!+\!1\!-\!2h)}
\prod_{j=0}^{\left[\frac{l\!-\!1}2\right]}\frac{1+\frac{(l\!-\!2j)^2}k}{1+\frac{(l\!-\!1\!-\!2j)^2}k}}\:\:;
\ea\label{transfD3}
\ee
here we have introduced the operator $\lambda:=[\sqrt{4L_iL_i+1}-1]/2$ (which has eigenvalues $l\in\{0,1,...,\Lambda\}$),
 $\Gamma$ is  Euler gamma function, the last equality holds only if $l\in\NN_0$, 
and $[b]$ stands for the integer part of $b$. 
 Therefore  the $L_{\lambda\mu }$  in the  $\bpi_{\Lambda}$-representation make up also an alternative set of generators of $\A_{\Lambda}$ (in \cite{FiorePisacaneJGP18}  $L_{4i}$ is denoted by $X_i$).

\item The group $Y_\Lambda\simeq SU(N)$ of $*$-automorphisms  of $\A_\Lambda$ is inner and includes a subgroup $SO(4)$ {\it independent of $\Lambda$}
(acting irreducibly via $\bpi_\Lambda$) and a subgroup $O(3)\subset SO(4)$ corresponding to orthogonal transformations (in particular, rotations) of the coordinates $x_i$, which play the role of isometries of $S^2_\Lambda$..

\item In the limit $\Lambda\to \infty$ \ 
dim$(\Hi_{\Lambda})\!\to\!\infty$,  \ and we recover QM on the sphere $S^2$ as sketched in section \ref{preliminari2'} (see \cite{FiorePisacaneJGP18} for details).

\end{itemize}

\subsection{$O(3)$-invariant  UR and strong SCS  on $S^2_\Lambda$}
\label{sphereCS1}

We first note that, since the commutation relations  $[L_i,L_j]=i\varepsilon^{ijk}  L_k $  
are as on $S^2$, then not only the UR  (\ref{LUR3}),  but also 
Theorem \ref{propLUR3'} and the resolution of the identity (\ref{ResolIdS^2})  hold, provided
$l$ runs over $\{0,1,..., \Lambda\}$ instead of $\NN_0$: 

\begin{teorema} The uncertainty relation 
\bea
(\Delta\bL)^2\ge |\langle \bL\rangle| \qquad\Leftrightarrow \qquad
\langle \bL^2\rangle
\ge |\langle \bL\rangle|\left(|\langle \bL\rangle|+1\right)  \label{LUR3''}
\eea
holds on $\Hi_\Lambda=\oplus_{l=0}^\Lambda V_l$ and is saturated by
the spin coherent states $\bm{\phi}_{l,g}:=  \bpi_\Lambda (g)\bpsi^l_l\in V_l$, $l\in \{0,1,..., \Lambda\}$, $g\in SO(3)$. Moreover on $\Hi_\Lambda$
the following resolution of identity holds:
\be
I=\sum_{l=0}^\Lambda C_l\int_{SO(3)} d\mu(g) P_{l,g},\qquad\qquad C_l=\frac{2l\!+\!1}{8\pi^2},\qquad  P_{l,g}=\bm{\phi}_{l,g}
\langle\bm{\phi}_{l,g},\cdot\rangle. 
\label{ResolIdS^2_L}
\ee
\label{propLUR3''}
\end{teorema}
\noindent
We can parametrize $g\in SO(3)$,  the invariant measure and the integral over $SO(3)$ through the Euler angles $\varphi,\theta,\psi$:  
\bea
&& g=e^{\varphi I_3}e^{\theta I_2}e^{\psi I_3}\quad\mbox{where }\:
I_3:=\left(\!\!\!\ba{ccc} 0 & 1 & 0\\ -1 & 0 & 0 \\ 0 & 0 & 0\ea\!\!\right)\!,\quad
I_2:=\left(\!\!\ba{ccc} 0 & 0 & -1\\ 0 & 0 & 0 \\ 1 & 0 & 0\ea\!\!\right) 
\quad\Rightarrow\\
&& \bpi_\Lambda(g)=
e^{i\varphi L_3}e^{i\theta L_2}e^{i\psi L_3}, \quad
\int\limits_{SO(3)}\!\!\!\! d\mu(g)=\!\!\int\limits^{2\pi}_0\!\!d\varphi\!\!\int\limits^{\pi}_0\!\!d\theta\sin\theta\!\!\int\limits^{2\pi}_0\!\!d\psi=8\pi^2.   
\eea

Since  the commutation relations $[L_i,x_j]=i\varepsilon^{ijk} x_k $ hold also on $S^2_{\Lambda}$,
so do  the UR   (\ref{HUR3}). However we will not investigate whether they (or some alternative 
ones) can be saturated, because to our knowledge this is not known even for the commutative $S^2$.

In the commutative case the spacial uncertainties $\Delta x_1, \Delta x_2, \Delta x_3$ can be
 simultaneously as small as we wish, because $ [x_i,x_j]=0$.
In the fuzzy case even the Robertson UR 
$$
4\left(\Delta x_1\right)^2 \left(\Delta x_2\right)^2\ge  \left\langle L'_3\right\rangle^2+\langle x_1x_2+x_2x_1\rangle^2,
\qquad L'_3:=\left(\!\frac{I}{k}-K\widetilde{P}_{\Lambda}\!\right)L_3, 
$$
and its permutations, which follow from (\ref{xx}) and are slightly stronger than the 
Schr\"odinger UR, are not particularly stringent, in that the right-hand side vanishes on a large class of states\footnote{In fact, on the generic vector
$\bm{\chi}=\sum\limits_{l=0}^{\Lambda}\sum\limits_{m=-l}^{l}{\chi_l^m\psi_l^m}$ one finds
$\left\langle L_3'\right\rangle_{\bm{\chi}}=\sum\limits_{l=0}^{\Lambda-1}\sum\limits_{m=1}^{l}[|\chi_l^{-m}|^2\!-\!|\chi_l^{m}|^2]\frac mk
+\frac{1+\frac{\Lambda^2}{k}}{2\Lambda+1}\sum\limits_{m=1}^{\Lambda}
[|\chi_{\Lambda}^{-m}|^2\!-\!|\chi_{\Lambda}^m|^2]$, which vanishes e.g. if 
$|\chi_l^{-m}|\!=\!|\chi_l^{m}|$ for all $l,m$, and $\langle x_1x_2+x_2x_1\rangle=\langle x_+^2-x_-^2\rangle/2i$,
which vanishes if e.g. all $\chi_l^{m}\in\RR$, so that $\langle x_+^2\rangle$ is real.}, hence does not exclude that either $\Delta x_1,\Delta x_2$ or  $\Delta x_3$ vanish.
However,  we will see that they cannot vanish simultaneously,  
because $(\Delta\bx)^2$ is bounded from below (see section \ref{CS3}).
Summing the Schr\"odinger UR
$$
\frac {\left(\Delta x_1\right)^4\!+\!\left(\Delta x_2\right)^4}2\ge \left(\Delta x_1\right)^2\left(\Delta x_2\right)^2
\ge\frac {\langle L_3'\rangle ^2}4 \quad\Rightarrow\quad 
\frac {\left(\Delta x_1\right)^4\!+\!\left(\Delta x_2\right)^4}2+2\left(\Delta x_1\right)^2\left(\Delta x_2\right)^2\ge \frac 3{4}\langle L_3'\rangle ^2,
$$
and the ones with permuted indices $1,2,3$  we find the $O(3)$-invariant UR
\be
\left(\Delta\bx\right)^4\ge \frac 3{4} \langle \bL'\rangle^2.     \label{x^4L^2UR}
\ee
Note that on the eigenstates of $x_0,L_0\!\equiv\! L_3$, with $L_0=m$ it is $\langle L_\pm'\rangle=0$ and 
$|\langle \bL'\rangle|=|\langle L_3'\rangle|=|m|\left(\!1/k\!-\!K\langle \widetilde{P}_{\Lambda}\rangle\!\right)$; in particular for $m=0$ the right-hand side of (\ref{x^4L^2UR}) is zero.
We leave it for possible future investigation to determine  the states, if any, saturating 
the UR (\ref{x^4L^2UR}); clearly there can be no saturation on a state such that $\langle L_3'\rangle=0$, because as said $(\Delta\bx)^2$ has a positive minimum.

We now apply  (\ref{PereloCS}) adopting  as a  $G$ not 
$SO(4)$ (the largest $\Lambda$-independent subgroup of the group of $*$-automorphism of $\A_\Lambda$), but its subgroup $G=SO(3)$ with Lie algebra spanned by the $L_i$, \ and $T\!=\!\bpi_\Lambda$. \ 
By (\ref{azioneL}), $(\Hi_\Lambda,\bpi_\Lambda)$ is a {\it reducible} representation of $G$,
more precisely  the direct sum of the irreducible representations $(V_l,\pi_l)$, $l=0,...,\Lambda$;
therefore completeness and resolution of the identity are not automatic.
Fixed a normalized vector   $\bm{\omega}\in\Hi_\Lambda$, for $g\in G$ let 
\be
\bm{\omega}_g:=\bpi_\Lambda(g)\bm{\omega}
,\qquad\qquad P_g:=\bm{\omega}_g\langle\bm{\omega}_g,\cdot\rangle.
\ee
The system  $A:=\{\bm{\omega}_g\}_{g\in G}$ is complete
provided that for all $l$ there exists at least one $h$ such that $\omega_l^h\neq 0$ (then it is also overcomplete).
In appendix \ref{ProofResolIdS^2_Lgen} we prove  

\begin{teorema}
If \ $\bm{\omega}=\sum\limits_{l=0}^{\Lambda}\sum\limits_{h=-l}^{l}
\omega_l^h\bpsi_l^h$ \ fulfills
\be
\sum\limits_{h=-l}^{l}|\omega_l^h|^2  =\frac{2l\!+\!1}{(\Lambda+1)^2},
\qquad l\!=\!0,1,...,\Lambda,       \label{NormCond}
\ee
then the following resolution of the identity on $\Hi_\Lambda$ holds:
\bea
I=\frac{(\Lambda+1)^2}{8\pi^2}\int_{SO(3)}\!\! d\mu(g) \,P_g,\qquad
P_g:=\bomega_g \langle\bomega_g,\cdot\rangle,\qquad \bm{\omega}_g:=\bpi_\Lambda(g)\bm{\omega}.\label{ResolIdS^2_Lomegagen}
\eea
If $\omega_l^h=\omega_l^{-h}$ the strong SCS $\{\bomega_g\}_{g\in SO(3)}$  is fully $O(3)$-equivariant.
\label{ResolIdS^2_Lgen}
\end{teorema}

In particular, choosing $\bm{\omega}=\bm{\omega}^\beta:=\sum_{l=0}^{\Lambda}
\bpsi_l^l e^{i\beta_l}\sqrt{2l\!+\!1}/(\Lambda\!+\!1)$ 
we find a family of strong SCS $\{\bomega_g^\beta\}_{g\in SO(3)}$ and associated  resolutions of the identity parametrized by \ $\beta\equiv(\beta_0,...,\beta_\Lambda)\in(\RR/2\pi\ZZ)^{\Lambda\!+\!1}$.
In appendix \ref{Ciccio} we compute $(\Delta \bL)^2,(\Delta \bx)^2$ on this strong SCS; 
the first is independent of $\beta,g$, the second is minimal if $\beta=0$. 
Then they are given by
\bea
(\Delta \bL)^2=\frac{\Lambda(2\Lambda^3\!+\!32\Lambda^2\!+\!65\Lambda\!+\!36)}{36(\Lambda\!+\!1)^2}, \qquad\qquad
(\Delta \bx)^2<\frac{3}{\Lambda+1} .
\label{LXURomega}
\eea
We can construct a  strong SCS  with a larger $(\Delta \bL)^2$ and a smaller $(\Delta \bx)^2$.  Choosing
$\bm{\omega}=\bphi^\beta=\sum_{l=0}^{\Lambda}\bpsi_l^0 e^{i\beta_l}\sqrt{2l\!+\!1}
/(\Lambda\!+\!1)$ [this is suggested by the arguments following (\ref{x^4L^2UR}) and the ones of next subsection] we again find a family of strong SCS and associated  resolutions of the identity parametrized by \ $\beta\equiv(\beta_0,...,\beta_\Lambda)\in(\RR/2\pi\ZZ)^{\Lambda\!+\!1}$. This SCS is fully $O(3)$-equivariant.
Since $\bm{\phi}^\beta$ are eigenvectors of $L_3$ (actually with zero eigenvalue), the isotropy group
$H=\{e^{i \psi L_3}\, |\, \psi\in\RR\}\simeq SO(2)$  is nontrivial, and the
resolution of the identity holds also with the integral extended over just the coset space
\ $S^2\simeq SO(3)/SO(2)\ni g=e^{\varphi I_3}e^{i\theta I_2}$:
\bea
I=\frac{(\Lambda+1)^2}{4\pi}\!\!\int^{2\pi}_0\!\!\!\!\!\!d\varphi \!\!\int^{\pi}_0\!\!\!\!d\theta\,\sin\theta \: P_g^\beta,\qquad
P_g^\beta=\bphi_g^\beta \langle\bphi_g^\beta,\cdot\rangle,\qquad \bphi_g^\beta=\sum_{l=0}^{\Lambda}
\frac{e^{i\beta_l}\!\sqrt{2l\!+\!1}}{\Lambda\!+\!1}\bpi_\Lambda(g)\bpsi_l^0.\label{ResolIdS^2_Lphi}
\eea
In the appendix we compute $(\Delta \bL)^2,(\Delta \bx)^2$ on the SCS 
$\{\bphi_g^\beta\}_{g\in G}$; this is the
analog of the SCS  (\ref{IdResol}-\ref{utileb}). Again
$(\Delta \bx)^2$ is smallest if $\beta=0$. Correspondingly, we find 
\bea
(\Delta \bL)^2=\frac{\Lambda(\Lambda\!+\!2)}{2}, \qquad\qquad
(\Delta \bx)^2<\frac{1}{\Lambda+1}.\label{LXURphi}
\eea



\subsection{$O(3)$-invariant weak SCS on $S^2_\Lambda$ minimizing $(\Delta \bm{x})^2$}
\label{CS3}

As $(\Delta \bm{x})^2$ is $O(3)$-invariant, so is the set ${\cal W}^2$ of states  
on $S^2_\Lambda$  minimizing 
$(\Delta \bm{x})^2$. Arguing as in the introduction, one can  first look for the states 
$\underline{\bm{\chi}}\in{\cal W}^2$
on which $\langle x_0\rangle=|\langle \bx\rangle|$ [whence $\langle x_\pm\rangle=0$, $(\Delta \bm{x})^2=\langle \bx^2\rangle-\langle x_0\rangle^2$], and then recover the whole ${\cal W}^2$ as
${\cal W}^2=\{\underline{\bm{\chi}}_g:=\bpi_\Lambda(g)\underline{\bm{\chi}} \, |\, g\in SO(3)\}$.
Presumably it is not possible to determine the most localized state $\underline{\bm{\chi}}^2$ in closed form  for general $\Lambda$.
Since eq. (\ref{xx}) implies that $\bx^2\geq \frac{1}{2}$ on the $\bL^2=\Lambda(\Lambda+1)$ eigenspace and  $\bx^2=1+O(1/\Lambda^2)$ on the orthogonal complement,  $(\Delta \bm{x})^2=\langle \bx^2\rangle-\langle x_0\rangle^2$ on the eigenvector  $\widehat{\bm{\chi}}$ of $x_0$ with highest eigenvalue exceeds $(\Delta \bm{x})^2_{min}$ at most by  a term $O(1/\Lambda^2)$. Presumably  it is not possible to determine $\widehat{\bm{\chi}}$ in closed form  for general $\Lambda$ either; determining analytically the eigenvalues and eigenvectors  of a square matrix of large rank is an absolutely nontrivial problem. Nevertheless in \cite{newpaper} we succeed in  carrying out a detailed study of their properties.
In particular, since $\left[x_0,L_0\right]=0$, we can simultaneously diagonalize $x_0$ and $L_0$. By (\ref{azioneL}) the eigenvalues of $L_0$ are $m\in\{-\Lambda,1\!-\!\Lambda,...,\Lambda\}$; 
let $\Hi_\Lambda^m$ be the corresponding eigenspaces. We  look for eigenvectors of both $x_0,L_0$ in the form
\be \label{diagxL}
\begin{cases}
L_0\bm{\chi}_{\alpha}^m=m \bm{\chi}_{\alpha}^m\\
x_0\bm{\chi}_{\alpha}^m=\alpha \bm{\chi}_{\alpha}^m
\end{cases}\!\!\!\!\!\!,\qquad\qquad 
\qquad\bm{\chi}_{\alpha}^m=\sum_{l=|m|}^{\Lambda}{\chi_l^m\bm{\psi}_l^m}.
\ee
Note that $L_0\bm{\chi}=m \bm{\chi}$ (with any $m$) implies $\left\langle x_\pm\right\rangle_{\bm{\chi}}=0$, $|\left\langle \bx\right\rangle_{\bm{\chi}}|=|\left\langle x_0\right\rangle_{\bm{\chi}}|$.
The second equation in  (\ref{diagxL}) turns 
out to be an eigenvalue equation for a real, symmetric and tri-diagonal square matrix $B_m(\Lambda)$ having dimension $\Lambda-|m|+1$. It is easy to see that we can restrict our attention to the cases $m\in\{0,1,\cdots,\Lambda\}$; in theorem 4.1 in \cite{newpaper} we prove that

\begin{enumerate}
\item If $\alpha$ is an eigenvalue of $B_m\left(\Lambda\right)$, then also $-\alpha$ is.

\item For all $\Lambda,m$, all  eigenvalues of $B_m\left(\Lambda\right)$ are simple; we denote them as \ $\alpha_1(\Lambda;m),
\alpha_2(\Lambda;m),$ $...,\alpha_{n(m)}(\Lambda;m)$, \ in decreasing order. The highest eigenvalues  $\alpha_1\left(\Lambda;m\right)$ of the $B_m\left(\Lambda\right)$ fulfill
\be\label{disalpha}
\alpha_1\left(\Lambda;0\right)>\alpha_1\left(\Lambda;1\right)>\cdots>\alpha_1\left(\Lambda;\Lambda\right),
\ee
\be 
\alpha_1\left(\Lambda+1;0\right)>\alpha_1\left(\Lambda;0\right)\quad\mbox{definitively, if }k(\Lambda)\geq\Lambda^6. 
\ee
\item The spectrum of $B_0$ becomes uniformly dense in $[-1,1]$ as $\Lambda\to \infty$.

\end{enumerate}

By (\ref{disalpha}), the eigenvector $\widehat{\bm{\chi}}$ of $x_0$ with the highest eigenvalue $\alpha_1\left(\Lambda;0\right)$ belongs to $\Hi_\Lambda^0$.
The matrix representing $x_0$ in the basis $\{\bpsi_l^0\}_{l=0,...,\Lambda}$ of  $\Hi_\Lambda^0$ is \cite{newpaper}
\be
B_0=B_0\left(\Lambda\right)=\left(
\begin{array}{cccccccc}
0&a_1&0&0&0&0&0&0\\
a_1&0&a_2&0&0&0&0&0\\
0&a_2&0&a_3&0&0&0&0\\
\vdots&\vdots&\vdots&\vdots&\vdots&\vdots&\vdots&\vdots\\
0&0&0&0&0&a_{\Lambda-1}&0&a_{\Lambda}\\
0&0&0&0&0&0&a_{\Lambda}&0\\
\end{array}
\right),\label{matral}
\ee
where
$$
a_l:=c_lA_l^{0,0}=\sqrt{1+\frac{l^2}{k}}\sqrt{\frac{l^2}{4l^2-1}}>\frac{1}{2}\qquad\forall l\leq \Lambda, \quad\forall\Lambda\in\mathbb{N},
$$
and this implies (see proposition 6.2 in \cite{newpaper})
\be
\left\|B_0\chi\right\|_2>\left\|\frac 12 P_{\Lambda+1}(0,1,1)\chi\right\|_2 \quad \forall\chi\in\mathbb{R}^{\Lambda+1}_+.
\ee
The normalized vector $\widetilde{\chi}\equiv(\widetilde{\chi}_0,...,\widetilde{\chi}_l)\in\mathbb{R}^{\Lambda+1}_+$ maximizing the right-hand side is the eigenvector of $\frac 12 P_{\Lambda+1}(0,1,1)$ with highest eigenvalue $\lambda_1=\cos[\pi/(\Lambda\!+\!2)]$:
$$
\widetilde{\chi}_l=\sqrt{\frac{2}{\Lambda+2}}\,\sin{\left[\frac{(l+1)\pi}{\Lambda+2} \right]},\qquad\qquad 0\leq l\leq \Lambda;
$$
Hence as the highest lower bound for  \ $\left\vert\left\langle \bx\right\rangle_{\widehat{\chi}}\right\vert=\left\langle x_0\right\rangle_{\widehat{\chi}}=\alpha_1\left(\Lambda;0\right)=\left\|B_0\widehat{\chi}\right\|_2/\left\|\widehat{\chi}\right\|$  \ we  find
\be
\alpha_1\left(\Lambda;0\right)\ge \left\langle x_0\right\rangle_{\widetilde{\chi}}=\left\|B_0\widetilde{\chi}\right\|_2>\left\|P_{\Lambda+1}\left(0,\frac{1}{2},\frac{1}{2}\right)\widetilde{\chi}\right\|_2=\cos\left(\frac{\pi}{\Lambda\!+\!2}\right).
\label{disscalprod}
\ee
This finally suggests that a quite stringent upper
 bound for $(\Delta \bm{x})^2_{min}$
should be $(\Delta \bm{x})^2$ on  $\widetilde{\bm{\chi}}=\sum_{l=0}^{\Lambda}\widetilde{\chi}_l\bm{\psi}_l^0\in\Hi_\Lambda^0$. In fact, in the appendix we show that
\bea\label{uncfinalenostra}
(\Delta \bm{x})^2_{\widetilde{\bm{\chi}}}\overset{\Lambda\geq 3}<
\frac{\pi^2}{(\Lambda+2)^2}+\frac{1}{(\Lambda+1)^2}<\frac{11}{(\Lambda+1)^2}.
\eea
This leads to the  important result  mentioned in the introduction:  the smallest space dispersion on our fuzzy sphere is smaller than the one (\ref{uncMadore})  on  the Madore's FS when $l=\Lambda$, i.e. the cutoffs of the two fuzzy spaces are the same; in formulas,
\be\label{confrontoMadVSNoi}
(\Delta \bm{x})^2_{min}\leq (\Delta \bm{x})^2_{\widetilde{\bm{\chi}}}<(\Delta \bm{x})^2_{min Madore}
\equiv \frac 1{\Lambda+1}.
\ee
Replacing $\underline{\bm{\chi}}$ by $\widehat{\bm{\chi}},\widetilde{\bm{\chi}}$ in
the definition of ${\cal W}^2$ we respectively obtain fully $O(3)$-invariant weak SCS  $\widehat{{\cal W}}^2,\widetilde{{\cal W}}^2$ approximating ${\cal W}^2$. 
Since $\widehat{\bm{\chi}},\widetilde{\bm{\chi}}$
are eigenvectors of $L_0$, the corresponding isotropy subgroup of $SO(3)$ is isomorphic to $SO(2)$, 
and the rays of the elements of $\widehat{\bm{\chi}},\widetilde{\bm{\chi}}$
are in one-to-one correspondence with the points of the sphere $S^2\simeq SO(3)/SO(2)$. 
The fact that the eigenvalue is zero  is in agreement with the classical picture of the particle:
the angular momentum $\bL=\bm{r}\wedge\bm{p}$ is orthogonal to the position vector $\bm{r}$,
hence if $\bm{r}\simeq \bm{k}$ (i.e. the particle is located concentrated around the north pole) then  $\bL$ is approximately orthogonal to the $x_3\equiv x_0$-axis,
and $L_3\equiv L_0\simeq 0$.

\section{Outlook, final remarks and conclusions}\label{Conclu}

In this paper we have introduced various strong and weak systems of coherent states  (SCS)\footnote{A strong SCS yields a resolution of the identity; 
a weak SCS is just (over)complete.} on  the fuzzy spheres $S^1_\Lambda,S^2_\Lambda$ 
and  studied their localizations in configuration as well as (angular) momentum space. As on the commutative spheres  $S^d$ ($d\!=\!1,2$), these localizations can be respectively expressed in terms of the uncertainties $\Delta x_i,\Delta L_{ij}$, or in terms  of their $O(D)$-invariant ($D\!\equiv\!d\!+\!1$) quadratic polynomials $(\Delta\bx)^2,(\Delta\bL)^2$  (sums of the variances of the $x_i$ and $ L_{ij}$, respectively); we have argued that the localizations expressed through
 $(\Delta\bx)^2,(\Delta\bL)^2$ are preferable because reference-frame independent. We have also determined general bounds (e.g. uncertainty relations following from commutation relations) for  $\Delta x_i,\Delta L_{ij},(\Delta\bx)^2,(\Delta\bL)^2$, estimated the latter on these SCS, partly investigated which SCS may saturate these bounds. Preliminarly we have discussed these issues for the commutative circle $S^1$ and sphere $S^2$, because the literature for the latter seems incomplete.

In particular we have derived the $O(3)$-invariant 
uncertainty relation (\ref{LUR3'}) (both on $S^2$ and  on $S_{\Lambda}^2$),  discussed its virtues compared  to the  $\Delta L_i\Delta L_j$ uncertainty relations (\ref{LUR3}), shown that the system of spin coherent states saturates it (see theorems \ref{propLUR3'} and \ref{propLUR3''}); also for the commutative $S^2$ this result is new. We have then discussed the Heisenberg (i.e. $\Delta x\Delta L$)  type uncertainty relations (HUR)  (\ref{HURS^1}),  which hold both on  $S^1$ and on $S_{\Lambda}^1$, and the states saturating them: we have shown that only the eigenvectors\footnote {The $\psi_n$  make an orthonormal basis of the Hilbert space; in a broad (but rather unconventional) sense this basis can be considered the system of coherent states associated to
the group (\ref{G1S^1}), semidirect product of a Lie group times a discrete one.} $\psi_n$ of $L$ saturate both (\ref{HURS^1})$_{1-2}$, or equivalently the $O(2)$-invariant
inequality (\ref{HURS^1})$_3$,  while
there is a complete family (parametrized by $\mu\in\RR$) of states saturating 
(\ref{HURS^1})$_1$ alone (these states are eigenvectors of $a_1^\mu:=L-i\mu x_1$); the family interpolates between the set of eigenvectors of $L$ and the set of eigenvectors of $x_1$. We have deferred an analogous discussion of HUR on $S_{\Lambda}^2$,
because the literature on this issue  on commutative $S^2$ seems even more incomplete.

Moreover,  for $d=1,2$ we have built a large class of strong SCSs on our fuzzy $S^d_\Lambda$ applying $SO(D)$-transformations on suitable initial states $\bomega\in\mathcal{H}_{\Lambda}$,
see eq. (\ref{IdResol}) and Theorem \ref{ResolIdS^2_Lgen}; 
in particular, we have chosen the SCS so as to minimize (within the class)
either  $(\Delta\bL)^2$, or $(\Delta\bx)^2$; the  SCS ${\cal S}^d$ minimizing $(\Delta\bx)^2$ is fully $O(D)$-equivariant, its states (rays)  are actually in one-to-one correspondence with points of $S^d\simeq SO(D)/SO(d)$, and their
$(\Delta\bx)^2$  is smaller than the uncertainty (\ref{uncMadore}) in Madore FS, i.e. satisfies  $(\Delta\bx)^2< 1/(\Lambda+1)$ - see (\ref{utileb}), (\ref{LXURphi}) [more careful computations will lead to lower upper bounds for $(\Delta\bx)^2$].

For both $d=1,2$ we have also introduced a fully $O(D)$-equivariant, weak SCS 
${\cal W}^d=\{\underline{\bm{\chi}}_g:=\pi_{\Lambda}(g)\underline{\bm{\chi}} \, |\, g\in SO(D)\}$ consisting of states minimizing  $(\Delta \bm{x})^2$ within the whole 
Hilbert space $\Hi_\Lambda$; the states (rays) of  ${\cal W}^d$ are actually
again  in one-to-one correspondence with the points of $S^d\simeq SO(D)/SO(d)$.
We have determined  them up to order $O(1/\Lambda^2)$, with the help of
the results of \cite{newpaper}: we have approximated 
the vector $\underline{\bm{\chi}}$ as the eigenvector $\widetilde{\bm{\chi}}$ with maximal eigenvalue of a suitable Toeplitz tridiagonal matrix, and denoted as $\widetilde{{\cal W}}^d$ the corresponding SCS; this eigenvector is  in turn  very close to the eigenvector with maximal eigenvalue of $x_1$ (resp. $x_0\equiv x_3$), because numerical computations suggest us that $\left\|X^{\Lambda}\right\|_2$ and $\left\|B_0(\Lambda)\right\|_2$ both converge with order $2$ to $1$.

For these reasons the strong SCS ${\cal S}^d$ (or alternatively the weak one ${\cal W}^d$, if we do not need a resolution of the identity) can be considered the system of quantum states
that is the ``closest" approximation to $S^d$.

We emphasize that the states of  the strong SCS ${\cal S}^2$ (resp. of  the weak SCS ${\cal W}^2$, $\widetilde{{\cal W}}^2$) are better localized than the most localized states of the Madore fuzzy sphere with the same cutoff ($l=\Lambda$)  by a factor smaller than 1, see (\ref{LXURphi}) [resp. by a power of $1/\Lambda$, see (\ref{confrontoMadVSNoi})]. On  $S^2_\Lambda$  the state $\bm{\chi}\in{\cal S}^2$ centered around  the North pole  (i.e. with 
$\langle x_1\rangle=\langle x_2\rangle=0$, $\langle x_3\rangle>0$)
fulfills the property $L_3\bm{\chi}=0$; the classical counterpart of this property is that a classical particle at the North pole of $S^2$  has zero $L_3$ ($z$-component of the angular momentum), see section \ref{CS3}. As noted in \cite{newpaper}, such a property is impossible to realize on the Madore-Hoppe
FS. For these reasons, and the other ones mentioned in the introduction,  we believe that
our fuzzy sphere $S^2_\Lambda$ is a much more realistic fuzzy approximation 
of a classical $S^2$ configuration space. 

Finally, the construction of various systems of coherent states on our fuzzy circle and fuzzy sphere will be very useful
to study quantum mechanics and above all quantum field theory on these fuzzy spaces.

\subsubsection*{Acknowledgments}

We are grateful to F. D'Andrea and F. Bagarello for useful discussions.

\section{Appendix}\label{Appendix}

\subsection{Proof of Theorems \ref{propLUR3'} and \ref{propLUR3''}}\label{proofpropoURS2}

Consider a generic Hilbert space $\Hi$ carrying a unitary representation of $O(3)$. For any vector $\psi\in \Hi$, let $g\in O(3)$ be a $3\times 3$ matrix such that 
 the expectation values of $L_j$ on $\psi$ fulfill
\be
g_{ij}\langle L_j\rangle=\delta_i^3|\langle \bL\rangle|.           \label{eq1}
\ee
 The expectation values of the $L_j,\bL^2$ on the states $\psi$, $\psi':=U(g)\psi$ fulfill $\langle L_1\rangle'=\langle L_2\rangle'=0$,
$\langle L_3\rangle'=|\langle \bL\rangle'|=|\langle \bL\rangle|\ge 0$, \ 
$\langle \bL^2\rangle'=\langle \bL^2\rangle$
(the second equalities hold because $U(g)$ is unitary). Hence
$\psi$ fulfills/saturates (\ref{LUR3'}) iff $\psi'$ respectively fulfills/saturates 
\be
\langle \bL^2\rangle' - \langle L_3\rangle'\left(\langle L_3\rangle'+1\right)\ge 0.
\label{ineq1}
\ee

If $\Hi=V_l$ the first term 
equals $l(l\!+\!1)$,  the inequality (\ref{ineq1}) is fulfilled,  and it is saturated  by $\psi'=|l,l\rangle$,  because \ Spec$(L_3)=\{-l,1\!-\!l,...,l\}$.

Now  assume that $\Hi$ can be decomposed as the direct sum
$\Hi=\Hi_1\oplus\Hi_2$ of orthogonal subspaces $\Hi_1,\Hi_2$ carrying subrepresentations of $O(3)$ 
 and on which  (\ref{LUR3'}) is fulfilled; moreover, let $\Gamma_i\subset\Hi_i$ 
be the subsets of vectors saturating (\ref{LUR3'}). Decomposing $\psi'=a_1\psi_1+a_2\psi_2$ and setting 
$\alpha:=|a_1|^2$, we  find $0\le\alpha\le 1$, $|a_2|^2=1\!-\!\alpha$, and
\bea
&&\langle \bL^2\rangle'-\langle L_3\rangle'\left(\langle L_3\rangle'+1\right) \label{eq2}\\[8pt]
&&=\alpha\langle \bL^2\rangle_1+(1\!-\!\alpha)\langle \bL^2\rangle_2
 -\left[\alpha\langle L_3\rangle_1+(1\!-\!\alpha)\langle L_3\rangle_2\right]^2
-\left[\alpha\langle L_3\rangle_1+(1\!-\!\alpha)\langle L_3\rangle_2\right]=:f(\alpha), \nonumber 
\eea
where we have abbreviated $\langle A\rangle_i\equiv \langle A\rangle_{\psi_i}$.
The polynomial $f'(\alpha)$  vanishes only at one point $\alpha'\in\RR$,
which however is of maximum for $f(\alpha)$, because $f''(\alpha)\!=\!-\left[\langle L_3\rangle_1\!-\!\langle L_3\rangle_2\right]^2\!\le\! 0$. Hence the minimum point of $f(\alpha)$ in the interval 
$[0,1]$ is either 0 or 1. But, by our assumptions,
\bea
f(1)=\langle \bL^2\rangle_1-|\langle \bL\rangle_1|\left(|\langle \bL\rangle_1|+1\right)\ge 0,\nn[6pt]
f(0)=\langle \bL^2\rangle_2-|\langle \bL\rangle_2|\left(|\langle \bL\rangle_2|+1\right)\ge 0, \nonumber
\eea
proving that (\ref{LUR3'}) is fulfilled on $\Hi$. Moreover, the set of states of $\Hi$
saturating the inequality is clearly $\Gamma=\Gamma_1\cup\Gamma_2$.

Choosing first $\Hi_1=V_0$ and $\Hi_2=V_1$, then $\Hi_1=V_0\oplus V_1$ and $\Hi_2=V_2$, and so on, one thus iteratively proves the statements of Theorems \ref{propLUR3'} and \ref{propLUR3''} for pure states.

Similarly we show that also mixed states (i.e. density operators) $\rho$  fulfill (\ref{LUR3'}), but cannot saturate it: abbreviating $\langle A\rangle\equiv \langle A\rangle_\rho:=\mbox{tr}(\rho A)$, let $g\in O(3)$ be a $3\times 3$ matrix such that 
 the expectation values of $L_j$ on $\rho$ fulfill (\ref{eq1}).
Then  the expectation values of $L_j,\bL^2$ on the state $\rho'=U(g)\rho U^{-1}(g)$ fulfill $\langle L_1\rangle'=\langle L_2\rangle'=0$,
$\langle L_3\rangle'=|\langle \bL\rangle'|=|\langle \bL\rangle|\ge 0$, \ 
$\langle \bL^2\rangle'=\langle \bL^2\rangle$, and
$\rho$ fulfills/saturates (\ref{LUR3'}) iff $\rho'$ fulfills/saturates 
(\ref{ineq1}). If $\rho'=\alpha\rho_1+(1\!-\!\alpha)\rho_2$,  the left-hand side of
(\ref{ineq1}) again takes the form (\ref{eq2}). Hence, reasoning as before, we find
that $\rho$ fulfills   (\ref{LUR3'}), and that there are no mixed states saturating this inequality.

\subsection{Some useful summations}\label{summations}

From $h(h\!+\!1)(h\!+\!2)...(h\!+\!j\!+\!1)-(h\!-\!1)h(h\!+\!1)...(h\!+\!j)=(j\!+\!2)h(h\!+\!1)...(h\!+\!j)$ (with $j\in\NN_0$) it follows
\be\label{sum_generic}
\sum\limits_{h=1}^n h(h\!+\!1)...(h\!+\!j)=\frac 1{j\!+\!2} n(n\!+\!1)(n\!+\!2)...(n\!+\!j\!+\!1); 
\ee
this implies, in particular,
\bea
\sum\limits_{h=1}^n h^2=\sum\limits_{h=1}^n [h(h\!+\!1)-h]=\frac { n(n\!+\!1)(n\!+\!2)}3-
\frac { n(n\!+\!1)}2=\frac { n(n\!+\!1)(2n\!+\!1)}6,\qquad \qquad \label{utile1} 
\eea
and in the following lines we will also use
\be \label{us_sumh_1}
\sum_{h=1}^{n}h^3=\frac{n^2(n+1)^2}{4},\quad\quad \sum_{h=1}^{n}h(2h+1)=\frac{4n^3+9n^2+5n}{6},
\ee
\be \label{us_sumh_2}
\sum_{h=1}^{n}h(h+1)(2h+1)=\frac{1}{2}n(n+1)^2(n+2),
\ee
\be \label{us_sumh_3}
\sum_{h=1}^{n}\left[h(h+1)+1\right](2h+1)=\frac{(n+1)^2(n^2+2n+2)}{2},\quad\quad \sum_{h=1}^{n}h\left(1-\frac{1}{2h}\right)=\frac{n^2}{2}.
\ee
Using the inequalities $1\!+\!x/2\ge\sqrt{1\!+\!x}\ge1\!+\!x/2\!-\!x^2/8$ (the first one is valid for
$x\ge -1$, the second for $x\le 8$) we  find
\bea
 1\!+\!\frac{m(m\!-\!1)}{2k}\ge b_m\ge  1\!+\!\frac{m(m\!-\!1)}{2k} \!-\!\frac{m^2(m\!-\!1)^2}{(2k)^2}\qquad   \label{utile3'}\\
\Rightarrow \qquad  n\!+\!\frac{(n\!-\!1)n(n\!+\!1)}{6k}\ge \sum\limits_{m=1}^{n} b_m\ge  n+\frac{(n\!-\!1)n(n\!+\!1)}{6k} -\frac{(n\!-\!1)n(n\!+\!1)(3n^2\!-\!2)}{60k^2}.                                      \label{utile3}
\eea
Using trigonometric formulae it is straightforward to show that
\be
\sum\limits_{m=2}^{n}\cos\left[\frac{\pi(2m\!-\!1)}{2n\!+\!2}\right]=0  \label{utile4}
\ee
(the terms cancel pairwise: the terms with $m=2,n$ cancel each other,  the terms with $m=3,n\!-\!1$  cancel each other, etc.),
and 
\bea
2\sin\!\left[\frac{\pi(n\!+\!1\!+\!m)}{2n\!+\!2}\right]\!\sin\!\left[\frac{\pi(n\!+\!m)}{2n\!+\!2}\right]\!&\!=\!&\!
\cos\!\left[\frac{\pi}{2n\!+\!2}\right]\!\!-\! \cos\!\left[\frac{\pi(2n\!+\!1\!+\!2m)}{2n\!+\!2}\right] \nn
\!&\!=\!&\! \cos\!\left[\frac{\pi}{2n\!+\!2}\right]\!+\cos\!\left[\frac{\pi(2m\!-\!1)}{2n\!+\!2}\right].\label{utile6}
\eea

\subsection{Proofs of some results regarding $S^1_\Lambda$}
\label{Lulu}

On a vector
$\bm{\chi}=\sum_{m=-\Lambda}^{\Lambda}{\chi_m\psi_m}$ we find \
$x_+\bm{\chi}=\sum_{m=-\Lambda}^{\Lambda-1}\chi_mb_{m+1}\psi_{m+1}$, and
\bea
\langle x_+\rangle_{\bm{\chi}} &=&
 \sum\limits_{m=1-\Lambda}^{\Lambda}\!\!\!\overline{\chi_m}\chi_{m-1}b_m; \label{utile1'}\\
\langle\bx^2\rangle_{\bm{\chi}} &=&  \sum\limits_{m=1-\Lambda}^{\Lambda-1}\left(1+\frac{m^2}{k}\right)|\chi_m|^2+ \frac 12\left[1+\frac{\Lambda(\Lambda-1)}{k}\right]\, 
(|\chi_{\Lambda}|^2+|\chi_{-\Lambda}|^2) 
\nn &=&\langle\bm{\chi},\bm{\chi}\rangle+  \sum\limits_{m=1-\Lambda}^{\Lambda-1}\frac{m^2}{k}
|\chi_m|^2+ \frac 12\left[\frac{\Lambda(\Lambda-1)}{k}-1\right]\, (|\chi_{\Lambda}|^2+|\chi_{-\Lambda}|^2) .
   \label{utile0}
\eea
We first prove (\ref{Lphiab}),
$$
\left\langle L\right\rangle_{\bm{\phi}_{\alpha}^{\beta}}=\frac{1}{2\Lambda+1}\sum_{m=-\Lambda}^{\Lambda}m=0,\quad \left\langle L^2\right\rangle_{\bm{\phi}_{\alpha}^{\beta}}=\frac{1}{2\Lambda+1}\sum_{m=-\Lambda}^{\Lambda}m^2=\frac{2}{2\Lambda+1}\sum_{m=1}^{\Lambda}m^2\overset{(\ref{utile1})}=\frac{\Lambda(\Lambda+1)}{3}.
$$
Now we prove (\ref{utile2'}). \ By (\ref{defR2D=2}), (\ref{utile1}), (\ref{utile1'}-\ref{utile0}) 
\bea
\langle\bx^2\rangle_{\bm{\phi}_\alpha^\beta} &=& \langle\bm{\phi}_\alpha^\beta,\bx^2\bm{\phi}_\alpha^\beta\rangle 
= 1+ \frac 2{2\Lambda\!+\!1}\sum\limits_{m=1}^{\Lambda}\frac{m^2}{k}-
 \frac 1{2\Lambda\!+\!1} \left[\frac{\Lambda(\Lambda+1)}{k}+1\right]\nn
&=& 1+ \frac {\Lambda(\Lambda\!+\!1)}{3k}-
 \frac 1{2\Lambda\!+\!1} \left[\frac{\Lambda(\Lambda+1)}{k}+1\right]
=\frac {2\Lambda }{2\Lambda\!+\!1}+ \frac {2(\Lambda\!-\!1)\Lambda(\Lambda\!+\!1)}{3(2\Lambda\!+\!1)k},  \nonumber 
 \eea
as claimed. Now we are able to prove (\ref{utileb}):
\bea
\left(\Delta\bx\right) ^2_{\bm{\phi}_\alpha^\beta} & =&\langle\bx^2\rangle_{\bm{\phi}_\alpha^\beta}-|\langle x_+\rangle_{\bm{\phi}_\alpha^\beta}|^2
=\frac {2\Lambda }{2\Lambda\!+\!1}+ \frac {2(\Lambda^2\!-\!1)}{3(2\Lambda\!+\!1)k}-\frac {4}{(2\Lambda\!+\!1)^2}\left[\sum\limits_{m=1}^{\Lambda} b_m\right]^2\nn
&\overset{1\le b_m}\leq &\frac {2\Lambda }{2\Lambda\!+\!1}+ \frac {2(\Lambda^2\!-\!1)\Lambda}{3(2\Lambda\!+\!1)k}-\frac {4\Lambda^2}{(2\Lambda\!+\!1)^2}\nn
&\overset{(\ref{kineq})}\leq & \frac {2\Lambda }{2\Lambda\!+\!1}-\frac {4\Lambda^2}{(2\Lambda\!+\!1)^2}+ \frac {2(\Lambda\!-\!1)}{3(2\Lambda\!+\!1)\Lambda(\Lambda+1)}
 <  \frac{2\Lambda}{(2\Lambda+1)^2}+\frac{1}{3\Lambda(\Lambda+1)}\nn
&<& \frac{2\Lambda}{4\Lambda(\Lambda+1)}+\frac{1}{3\Lambda(\Lambda+1)}= \frac 1{\Lambda+1}
\left(\frac 12+\frac{1}{3\Lambda}\right)\overset{\Lambda\ge 2}\le  \frac 2{3(\Lambda+1)}.
\nonumber
\eea

\medskip
We now prove (\ref{underlinechi}).
On a generic normalized $\bm{\chi}$ (\ref{utile1'}-\ref{utile0}) with $\Lambda=1$ gives 
\bea
\langle \bx^2\rangle_{\bm{\chi}} =\frac 12\left[1+|\chi_0|^2\right]=\frac 12\left[1+s\right],\qquad
\langle x_+\rangle_{\bm{\chi}} =\overline{\chi_0}\chi_{-1}+\overline{\chi_1}\chi_0,\nn
|\langle x_+\rangle_{\bm{\chi}}|^2=|\chi_0|^2\left(|\chi_1|^2\!+\!|\chi_{-1}|^2\right)+(\chi_0^2\overline{\chi_1}\overline{\chi_{-1}}\!+\!\overline{\chi_0}^2\chi_1\chi_{-1})=s(1-s)
+2st\cos\alpha,\nn
(\Delta\bx)^2_{\bm{\chi}} =\langle\bx^2\rangle_{\bm{\chi}}-|\langle x_+\rangle_{\bm{\chi}}|^2
=\frac 12\left[1-s\right]+s^2-2st\cos\alpha                  \label{temp} 
\eea
where $s:=|\chi_0|^2\le 1$, $t:=|\chi_1\chi_{-1}|$, and $\alpha$ is the phase of $\chi_0^2\overline{\chi_1}\overline{\chi_{-1}}$; by the Cauchy-Schwarz inequality $t\le\left(|\chi_1|^2\!+\!|\chi_{-1}|^2\right)/2=(1\!-\!s)/2$.
For fixed $s$, (\ref{temp}) is minimized by $\alpha=0$ and 
 $t=(1\!-\!s)/2$ (namely $|\chi_1|=|\chi_{-1}|=\sqrt{t}=\sqrt{(1\!-\!s)/2}$), what then yields
$$
(\Delta\bx)^2_{\bm{\chi}} =\frac 12(1\!-\!s)+s^2-s(1\!-\!s)=2s^2-\frac 32 s+\frac 12.
$$
This is minimized by $s=3/8$, and the minimum value is $(\Delta\bx)^2_{{min}} =7/32$, as claimed.
The corresponding minimizing vectors are $\underline{\bm{\chi}}=
\frac{\sqrt{5}}4\left[e^{i\beta}\psi_{-1}\!+\!e^{i\gamma}\psi_1\right]+\frac{\sqrt{3}}{\sqrt{8}}e^{i(\beta+\gamma)/2}\psi_0$; the one in (\ref{underlinechi}) is chosen so that $\langle x_+\rangle\in\RR$.

\bigskip
Next, we prove (\ref{Deltax2qminS^1_L}).
Up to normalization, the components of  the eigenvector $\bm{\chi}$ of the Toeplitz matrix $X_0$ with the maximal eigenvalue ($\lambda_M=\cos\left[\pi/(2\Lambda\!+\!2)\right]$)  are [see  (\ref{Toeplitz})]
\be
\chi_m=\sin\left[\frac{\pi(\Lambda\!+\!1\!+\!m)}{2\Lambda\!+\!2}\right]=
\cos\left[\frac{\pi m}{2\Lambda\!+\!2}\right];
\ee
then $\langle\bm{\chi},\bm{\chi}\rangle =\Lambda\!+\!1$,
\begin{equation*}
\begin{split}
 \langle\bm{\chi},\bx^2\bm{\chi}\rangle \stackrel {(\ref{utile0})}{=} &\langle\bm{\chi},\bm{\chi}\rangle+ 2\sum\limits_{m=1}^{\Lambda-1}\frac{m^2}{k}\chi_m^2+
 \left[\frac{\Lambda(\Lambda-1)}{k}-1\right]\, \chi_{\Lambda}^2 =  \Lambda+1+ 2\sum\limits_{m=1}^{\Lambda-1}\frac{m^2}{k}\chi_m^2+
 \left[\frac{\Lambda(\Lambda-1)}{k}-1\right]\, \chi_{\Lambda}^2\\
\overset{\chi_m^2\leq 1}\leq &  \Lambda+1+ 2\sum\limits_{m=1}^{\Lambda-1}\frac{m^2}{k}+
 \left[\frac{\Lambda(\Lambda-1)}{k}-1\right]\, \chi_{\Lambda}^2 \overset{(\ref{kineq})}\leq  \Lambda+1+ 2\sum\limits_{m=1}^{\Lambda-1}\frac{m^2}{k}\stackrel {(\ref{utile1})}{=} \Lambda+1+\frac{\Lambda (\Lambda-1)(2\Lambda-1)}{3k},
\end{split}
\end{equation*}
which implies
\be \label{ineqD=21}
\begin{split}
\langle\bx^2\rangle_{\bm{\chi}}=
\frac{\langle\bm{\chi},\bx^2\bm{\chi}\rangle}{\langle\bm{\chi},\bm{\chi}\rangle}\le & 1+ \frac{\Lambda(\Lambda-1)(2\Lambda-1)}{3k(\Lambda+1)}\overset{(\ref{kineq})}\leq 1+ \frac{\Lambda(\Lambda-1)(2\Lambda-1)}{3\Lambda^2(\Lambda+1)^3}\leq 1+ \frac{1}{(\Lambda+1)^2}.
\end{split}
\ee
Moreover,  due to (\ref{utile3'}),  (\ref{utile3}), $\chi_{-m}=\chi_m\in\RR$, it is $\langle x_1\rangle_{\bm{\chi}}=\langle x_+\rangle_{\bm{\chi}}$ because the latter is real, whence
\bea \label{ineqD=2,2}
\langle \bm{\chi},x_1\bm{\chi}\rangle &\stackrel{(\ref{utile1'})}{=}&  2 \sum\limits_{m=1}^{\Lambda}\! b_m\sin\left[\frac{\pi(\Lambda\!+\!1\!+\!m)}{2\Lambda\!+\!2}\right]\sin\left[\frac{\pi(\Lambda\!+\!m)}{2\Lambda\!+\!2}\right] \nn
&\overset{b_m\geq 1}\geq & 2 \sum\limits_{m=1}^{\Lambda}\! \sin\left[\frac{\pi(\Lambda\!+\!1\!+\!m)}{2\Lambda\!+\!2}\right]\sin\left[\frac{\pi(\Lambda\!+\!m)}{2\Lambda\!+\!2}\right] \nn
& \stackrel {(\ref{utile6})}{=} & \sum\limits_{m=1}^{\Lambda} \left\{\cos\left[\frac{\pi}{2\Lambda\!+\!2}\right]
+\cos\left[\frac{\pi(2m\!-\!1)}{2\Lambda\!+\!2}\right]\right\}\overset{(\ref{utile4})}=(\Lambda+1)\cos\left[\frac{\pi}{2\Lambda\!+\!2}\right]\quad\Longrightarrow\nn
\langle x_1\rangle_{\bm{\chi}}^2 =
\left(\!\frac{\langle \bm{\chi},x_1\bm{\chi}\rangle}{\langle\bm{\chi},\bm{\chi}\rangle}\!\right)^2 
&\ge &\cos^2\left[\frac{\pi}{2\Lambda\!+\!2}\right]=1-\sin^2\left[\frac{\pi}{2\Lambda\!+\!2}\right]\geq 1-\left(\frac{\pi}{2\Lambda\!+\!2}\right)^2,
\eea
\bea
\left(\Delta{\bx}\right)^2_{\bm{\chi}} &=&\langle\bx^2\rangle_{\bm{\chi}}- \!\langle x_1\rangle_{\bm{\chi}}^2  \nn &\overset{(\ref{ineqD=21})\&(\ref{ineqD=2,2})}\leq&
1+\frac{1}{(\Lambda+1)^2}-1+\left(\frac{\pi}{2\Lambda+2}\right)^2=\frac{1+\frac{\pi^2}{4}}{(\Lambda+1)^2}<\frac{3.5}{(\Lambda+1)^2}.
\eea

\subsection{States saturating the Heisenberg UR  (\ref{HURS^1}) on $S^1,S^1_\Lambda$}
\label{diverged=1}

For any $\mu\in\RR$, $i=1,2$
let 
$a^\mu_i:=L\!-\!i\mu x_i$, $z_i:=\langle L\rangle\!-\! i\mu\langle x_i\rangle$,  
$A^\mu_i:=
a^\mu_i-z_i$. \  
The inequality $0\le\langle A^\mu_i{}^\dagger A^\mu_i\rangle=\left(\Delta L\right)^2+\mu^2(\Delta x_i)^2+\mu\epsilon^{ij}\langle x_j\rangle$ 
(here $\epsilon^{11}\!=\!\epsilon^{22}\!=\!0$, $\epsilon^{12}\!=\!-\epsilon^{21}\!=\!1$, and a sum over $j\!=\!1,2$ is understood) is saturated on  the states 
annihilated by $A^\mu_i$, which are the eigenvectors 
$\bm{\chi}=\sum_{n}\chi_n\psi_n$ 
of $a^\mu_i$; here the sum runs over $n\in\ZZ$ for $S^1$ [where by $\psi_n$ we mean 
$(x_+)^n=e^{in\varphi}$], over $n\in I_\Lambda:=\{-\Lambda, 1\!-\!\Lambda,...,\Lambda\}$   for $S^1_\Lambda$. We can just stick to $i\!=\!1$; the UR will be thus saturated on the eigenvectors
of $a^\mu_1$.
The results for $a^\mu_2$ can be obtained by a rotation of $\pi/2$, by the $O(2)$-equivariance. 

One easily checks that $a^\mu_1\bm{\chi}=z{}\bm{\chi}$ in $\Hi={\cal L}^2(S^1)$ amounts to the equations
\be
2\chi_n(n-z{})-i\mu(\chi_{n+1}+\chi_{n-1})=0, \qquad n\in\ZZ.       \label{eqsn}
\ee
One way to fulfill them (with a non trivial $\bm{\chi}$) is with $\mu=0$; this implies $\chi_n=0$ for all $n$ but one, i.e. $\bm{\chi}\propto\psi_m$ for some $m\in\ZZ$, and $z{}=\langle L\rangle=m$. This is actually the only way: if $\mu\neq 0$ then 
the equations can be used as recurrence relations to determine all the $\chi_n$ as combinations of two, e.g. 
$\chi_0,\chi_1$; if the latter vanish so do all $\chi_n$, otherwise  the resulting sequence does not lead to a $\bm{\chi}\in\Hi$ because $\sum_n|\chi_n|^2=\infty$. In fact, rewriting (\ref{eqsn}) in the form $\chi_{n+1}=-\chi_{n-1}+C_n\chi_n$, with
$C_n:=\frac 2{i\mu}(n-z{})$ it is easy to  iteratively prove the relation
$$
\chi_{n+1}=\chi_{n}Q_n -\frac {\chi_0}{Q_1Q_2....Q_{n-1}},\qquad Q_1:=C_1,\quad Q_n:=C_n-\frac 1{Q_{n-1}}.
$$
This implies that as $n\to \infty$ $|C_n|\to\infty$, $|Q_n|\simeq|C_n|\to\infty$, 
$|\chi_{n+1}/\chi_{n}|^2\simeq|Q_n|^2\to\infty$, whence by the D'Alembert criterion the
series $\sum_{n=0}^\infty|\chi_{n}|^2$ diverges. The $\psi_m$ are also eigenvectors
of $a^2_{\mu=0}$ and therefore saturate not only 
(\ref{HURS^1})$_1$, but also (\ref{HURS^1})$_2$, and therefore all of (\ref{HURS^1}).

One easily checks that  the eigenvalue equation $a^\mu_1\bm{\chi}=z{}\bm{\chi}$ in $\Hi_\Lambda$ (i.e. on $S^1_\Lambda$)  amounts to the equations
\bea
\ba{ll}
2\chi_{-\Lambda}(\Lambda+z{})+i\mu \, b_{1-\Lambda}\chi_{1-\Lambda}=0,&\\[6pt]
2\chi_n(n-z{})-i\mu(b_{n+1}\chi_{n+1}+b_n\chi_{n-1})=0, \qquad & n=1\!-\!\Lambda,2\!-\!\Lambda,...,\Lambda\!-\!1,\\[6pt]
2\chi_{\Lambda}(\Lambda-z{})-i\mu \, b_{\Lambda}\chi_{\Lambda-1}=0 &
\ea                                                 \label{eqsn'}
\eea
(actually the second equations include also the first, third, because for $n\!=\!\pm\Lambda$, \ 
$b_{-\Lambda}\!=\!b_{\Lambda+1}=0$).
One way to fulfill (\ref{eqsn'})  
is with $\mu\!=\!0$; this implies $\chi_n\!=\!0$ for all $n$ but one, i.e. $\bm{\chi}\propto\psi_m$ for some $m\!\in\! I_\Lambda$, and $z{}\!=\!\langle L\rangle\!=\!m$. 
But   nontrivial solutions exist also with nonzero $\mu\!\neq\! 0$. In fact, equations (\ref{eqsn'})  can be used as recurrence relations to determine 
all the $\chi_n$ in terms of one. If we use them in the order to express first $\chi_{1-\Lambda}$ as $\chi_{-\Lambda}$ times a factor, then $\chi_{2-\Lambda}$ as $\chi_{-\Lambda}$ times another factor, etc., then the last equation amounts to the eigenvalue equation, a polynomial equation in $z$ of degree $(2\Lambda\!+\!1)$.
Note that if $z$ is an eigenvalue and $\bm{\chi}$ the corresponding eigenvector then
 also $z'\!=\!-z$ is an eigenvalue with corresponding eigenvector characterized by components $\chi_n'\!=\!(-1)^n\chi_{-n}$.
Since $a^\mu_2=e^{-i\pi L/2}a^\mu_1 e^{i\pi L/2}$, to each eigenvector  $\bm{\chi}$ of $a^\mu_1$ there corresponds the one $\bm{\chi}'=e^{-i\pi L/2}\bm{\chi}$ of $a^\mu_2$ with the same eigenvalue $z$ and components related by $\chi'_n=\chi_n(-i)^n$. Hence
$\bm{\chi}$ cannot be a simultaneous eigenvector of $a^\mu_1,a^\mu_2$ and therefore
again cannot saturate all of (\ref{HURS^1}), but only one of the first two inequalities,
unless $\mu=0$, namely unless it is an eigenvector of $L$; hence again the $\psi_m$ are
the only states  saturating all of (\ref{HURS^1}).

%

We determine the eigenvectors of $a_1^\mu$ for $\Lambda\!=\!1$. The eigenvalue equation amounts to
$z(z^2\!-\!1\!+\!\mu^2/2)\!=\!0$. We easily find that (\ref{eqsn'}) admits the following solutions:
\bea
z{}=0, \, \pm\sqrt{1\!-\!\frac{\mu^2}2} ,\qquad \bm{\chi}=\chi_{-1}
\left\{\psi_{-1}+
\frac{2i}{\mu}(1\!+\!z{})\psi_0-
\left[1\!+\!\frac{4z{}}{\mu^2}(1\!+\!z{})\right]\psi_1\right\}.
\eea
$\Vert\bm{\chi}\Vert^2\!=\!1$ amounts to
\ $\frac{|\chi_{-1}|^2}{\mu^4}\left\{\mu^4+4\mu^2|1\!+\!z{}|^2+
\left|\mu^2\!+\!4z{}(1\!+\!z{})\right|^2\right\}=1$. \ This leads to
\bea
&& z{}=0\qquad\Rightarrow\qquad |\chi_{-1}|^2=\frac{\mu^2}{2\mu^2+4} \nn
&& z{}=\pm\sqrt{1\!-\!\frac{\mu^2}2}\qquad\Rightarrow\qquad
z\in\left\{\!\!\ba{l}\RR,\\[6pt] i\RR,\ea\right. \quad |\chi_{-1}|^2
=\left\{\!\!\ba{ll}\frac{\mu^4}{32(1+z)-8\mu^2}\quad &\mbox{if }\mu^2\le 2,\\[6pt]
\frac 14\quad &\mbox{if }  \mu^2\ge 2.\ea\right. \nonumber
\eea
In the $\mu\!\to\!0$ limit we recover the eigenvectors $\psi_1,\psi_0,\psi_{-1}$ of $L$
with eigenvalues $-1,0,1$, whereas in the $\mu\!\to\!\infty$ limit we recover the eigenvectors 
$\varphi_-,\varphi_0,\varphi_+$ of $x_1$ with eigenvalues $-\sqrt{2}/2,0,\sqrt{2}/2$
(we obtain them in the reverse order $\varphi_+,\varphi_0,\varphi_-$ 
in the limit $\mu\!\to\!-\infty$). On the other hand if $\mu^2\!=\!2$ then all 
eigenvalues coincide with the zero eigenvalue, which remains with geometric multiplicity 1; in other words, in this case (only) there is no basis of $\Hi_\Lambda$ consisting of eigenvectors of $a_1^\mu$.  Moreover, recalling that  
$z=\langle L\rangle\!-\! i\mu\langle x_1\rangle$ we find that if $\mu^2\!\le\!2$ then 
$\langle x_1\rangle\!=\!0$ on all eigenvectors (because $z$ is real), whereas if $\mu^2\!\ge\!2$ then 
$\langle L\rangle\!=\!0$ on all eigenvectors  (because $z$ is purely imaginary).
One easily checks that 
$$
\langle x_1\rangle+ i\langle x_2\rangle=\langle x_+\rangle
=\frac{2i}{\mu}|\chi_{-1}|^2\left[2+z+\bar z+\frac{4z}{\mu^2}|1\!+\!z{}|^2\right],
$$
leading to
\bea
z{}=0,\quad \mu^2\le 2\qquad &\Rightarrow & \qquad \langle x_1\rangle=0,\quad\langle x_2\rangle=\frac {2\mu} {\mu^2\!+\!2},\quad\langle \bx\rangle^2=
\frac {4\mu^2} {(\mu^2\!+\!2)^2},\\[8pt]
z{}=0,\quad \mu^2\ge 2\qquad &\Rightarrow & \qquad \langle x_1\rangle=0,\quad\langle x_2\rangle=\frac 1{\mu},\quad\langle \bx\rangle^2=\frac 1{\mu^2},\\[8pt]
z{}=\pm\sqrt{1\!-\!\frac {\mu^2}2},\quad \mu^2\le 2\qquad &\Rightarrow & \qquad \langle x_1\rangle=0,
\quad\langle x_2\rangle=\frac {\mu}2,\quad\langle \bx\rangle^2=\frac {\mu^2}4,\\[8pt]
z{}=\pm i\sqrt{\frac {\mu^2}2\!-\! 1},\quad \mu^2\ge 2\qquad &\Rightarrow & \qquad \langle x_1\rangle=\frac {\mp 1}{\mu}\sqrt{\frac {\mu^2}2\!-\! 1},
\quad\langle x_2\rangle=\frac 1{\mu},\quad\langle \bx\rangle^2=\frac 12.
\eea

As on $\Hi_1$ it is $\bx^2=1-(\tilde P_1\!+\!\tilde P_{-1})/2$ we find
$$
\langle \bx^2\rangle
=1-\frac{|\chi_{-1}|^2}2\left\{1\!+\!\left|1\!+\!\frac{4z}{\mu^2}(1\!+\!z)\right|^2\right\}
$$
leading to
\bea
z{}=0,\quad \mu^2\le 2\qquad &\Rightarrow & \qquad \langle \bx^2\rangle=\frac {\mu^2\!+\!4} {2(\mu^2\!+\!2)},\quad(\Delta\bx)^2=\frac 12+\frac{2\!-\!3\mu^2}{(\mu^2\!+\!2)^2},
\qquad\\[8pt]
z{}=0,\quad \mu^2\ge 2\qquad &\Rightarrow & \qquad \langle\bx^2\rangle=\frac {\mu^2\!+\!4} {2(\mu^2\!+\!2)},\quad(\Delta\bx)^2=\frac {\mu^4\!+\!2\mu^2\!-\!4} {2\mu^2(\mu^2\!+\!2)},
\qquad\\[8pt]
z{}=\pm\sqrt{1\!-\!\frac {\mu^2}2},\quad \mu^2\le 2\qquad &\Rightarrow & \qquad \langle \bx^2\rangle=\frac 12\!+\!\frac{\mu^2}{8},
\quad(\Delta\bx)^2=\frac 12\!-\!\frac{\mu^2}{8},
\qquad\\[8pt]
z{}=\pm i\sqrt{\frac {\mu^2}2\!-\! 1},\quad \mu^2\ge 2\qquad &\Rightarrow & \qquad \langle \bx^2\rangle=\frac 34,
\quad(\Delta\bx)^2=\frac 14. \quad \qquad
\eea
We also find
\bea
z{}=0,\quad \qquad &\Rightarrow & \qquad \left(\Delta L\right)^2=\langle L^2\rangle=\frac {\mu^2} {\mu^2\!+\!2},\qquad\\[8pt]
z{}=\pm\sqrt{1\!-\!\frac {\mu^2}2},\quad \mu^2\le 2\qquad &\Rightarrow & \qquad \left(\Delta L\right)^2=1\!-\! \frac {\mu^2}4\!-\! \left[1\!-\!\frac{\mu^2(1\!+\!z)}{4(1\!+\!z)\!-\!\mu^2} \right]^2, \qquad\\[8pt]
z{}=\pm i\sqrt{\frac {\mu^2}2\!-\! 1},\quad \mu^2\ge 2\qquad &\Rightarrow & \qquad \left(\Delta L\right)^2=\langle L^2\rangle=\frac 12. \quad \qquad
\eea

For all $\mu$ $\bm{\chi}_\alpha:=e^{i\alpha L}\bm{\chi}$ is characterized by the same $(\Delta \bL)^2,(\Delta\bx)^2$ as  $\bm{\chi}$.
For all $\mu\neq 0$ and any of the eigenvectors $\bm{\chi}$ of $a^\mu_1$  
the system $X:=\{\bm{\chi}_\alpha\}_{\alpha\in[0,2\pi[}$ is complete
(actually overcomplete), but the resolution of the identity \ $\int_0^{2\pi}\!\! d\alpha \,\bm{\chi}_\alpha\langle\bm{\chi}_\alpha,\cdot\rangle=c I$ \ does not hold.

\subsection{Proof of Theorem \ref{ResolIdS^2_Lgen}}
\label{ProofResolIdS^2_Lgen}

This is based on the following two lemmas:
\begin{lemma}
Let $P^h\!=\!\sum_{l=|h|}^{\Lambda} \bpsi_l^h\langle\bpsi_l^h,\cdot\rangle$ be the projector on the
$L_3\!=\!h$ eigenspace. Then
\be
\int^{2\pi}_0\!\!\!\!d\alpha \: e^{i\alpha (L_3-h)}=2\pi P^h     \label{P^h}.
\ee
\end{lemma}
\noindent This can be proved applying both sides to the basis vectors $\bpsi_l^m$.
 In subsection \ref{prooflemmadeltaijpsi} we prove
 
\begin{lemma}\label{lemmadeltaijpsi}
If $\vert h\vert,\vert n\vert\leq l,j$ then
\be\label{deltaijpsi}
 \int^{\pi}_0\!\!\!\!d\theta\sin\theta \:
\langle\bpsi_{j}^n,e^{i\theta L_2}\bpsi_{j}^{h}\rangle\: \langle e^{i\theta L_2}\bpsi_l^{h},\bpsi_l^n\rangle=\frac{2}{2l+1}\delta_{lj}.
\ee
\end{lemma}
\noindent
Now let $B:=\int_{SO(3)}\!\! d\mu(g) \,P_g^\beta$, with a generic   $\bm{\omega}=\sum\limits_{l=0}^{\Lambda}\sum\limits_{h=-l}^{l}
\omega_l^h\bpsi_l^h$; we compute $B\bpsi_l^n$ ($|n|\le l$):
\bea
B\bpsi_l^n=\int^{2\pi}_0\!\!\!\!\!\!d\varphi\int^{\pi}_0\!\!\!\!d\theta\sin\theta\int^{2\pi}_0\!\!\!\!\!\!d\alpha
\: e^{i\varphi L_3}e^{i\theta L_2}e^{i\alpha L_3}\bm{\omega}\: \langle e^{i\theta L_2}e^{i\alpha L_3}\bm{\omega},e^{-i\varphi L_3}\bpsi_l^n\rangle \nn
\stackrel{(\ref{azioneL})}{=}\int^{2\pi}_0\!\!\!\!\!\!d\varphi\,e^{i\varphi (L_3-n)}\int^{\pi}_0\!\!\!\!d\theta\sin\theta\int^{2\pi}_0\!\!\!\!\!\!d\alpha
\:  e^{i\theta L_2}e^{i\alpha L_3}\bm{\omega}\: \langle e^{i\theta L_2}e^{i\alpha L_3}\bm{\omega},\bpsi_l^n\rangle \nn
\stackrel{(\ref{P^h})}{=}2\pi\sum_{j=|n|}^{\Lambda} \bpsi_{j}^n\int^{\pi}_0\!\!\!\!d\theta\sin\theta\int^{2\pi}_0\!\!\!\!\!\!d\alpha
\: \langle\bpsi_{j}^n,e^{i\theta L_2}e^{i\alpha L_3}\bm{\omega}\rangle\: \langle e^{i\theta L_2}e^{i\alpha L_3}\bm{\omega},\bpsi_l^n\rangle\nn
=2\pi\sum_{j=|n|}^{\Lambda} \bpsi_{j}^n\int^{\pi}_0\!\!\!\!d\theta\sin\theta\int^{2\pi}_0\!\!\!\!\!\!d\alpha
\: \sum_{h=-l}^{l}\overline{\omega_l^h}\sum_{m=-j}^{j}
\omega_j^m\langle\bpsi_{j}^n,e^{i\theta L_2}e^{i\alpha L_3}\bpsi_{j}^{m}\rangle\: \langle e^{i\theta L_2}e^{i\alpha L_3}\bpsi_l^{h},\bpsi_l^n\rangle
\nn
=2\pi\sum_{j=|n|}^{\Lambda} \bpsi_{j}^n\sum_{h=-l}^{l}\overline{\omega_l^h}\sum_{m=-j}^{j}\omega_j^m
\int^{\pi}_0\!\!\!\!d\theta\sin\theta\int^{2\pi}_0\!\!\!\!\!\!d\alpha \: e^{i\alpha (m-h)}
\langle\bpsi_{j}^n,e^{i\theta L_2}\bpsi_{j}^{m}\rangle\: \langle e^{i\theta L_2}\bpsi_l^{h},\bpsi_l^n\rangle
\nn
\stackrel{(\ref{P^h})}{=}(2\pi)^2\sum_{j=|n|}^{\Lambda} \bpsi_{j}^n\sum_{h=-m_{jl}}^{m_{jl}}\overline{\omega_l^h}\omega_j^h \int^{\pi}_0\!\!\!\!d\theta\sin\theta \:
\langle\bpsi_{j}^n,e^{i\theta L_2}\bpsi_{j}^{h}\rangle\: \langle e^{i\theta L_2}\bpsi_l^{h},\bpsi_l^n\rangle
\nonumber
\eea
where   $m_{jl}\!:=\!\min\{j,l\}$. \ By (\ref{deltaijpsi}) this becomes 
$B\bpsi_l^n=\bpsi_{l}^n\!\!\sum\limits_{h=-l}^{l}
|\omega_l^h|^2 8\pi^2/(2l+1)$. \
In order that this equals $C\bpsi_l^n$, i.e. that $B=CI$ with some constant $C\!>\!0$, it must be \
$\sum\limits_{h=-l}^{l}|\omega_l^h|^2  =C(2l\!+\!1)/8\pi^2$ \ for all $l\!=\!0,...,\Lambda$.
Summing over $l$ and imposing that $\bomega$ be normalized  we find
\be
1=\Vert\bm{\omega}\Vert^2=\sum_{l=0}^{\Lambda}\sum\limits_{h=-l}^{l}|\omega_l^h|^2
=\sum_{l=0}^{\Lambda}\frac{2l+1}{8\pi^2}C=\frac{(\Lambda+1)^2}{8\pi^2}C
\quad\Rightarrow\quad C=\frac{8\pi^2}{(\Lambda+1)^2},    \label{C=}
\ee
as claimed. The strong SCS $\{\bomega_g\}_{g\in SO(3)}$  is fully $O(3)$-equivariant if $\omega_l^h=\omega_l^{-h}$ , because then it is mapped 
into itself also by the unitary transformation $\bpsi_l^h\mapsto \bpsi_l^{-h}$ that corresponds to 
the transformation of the coordinates (with determinant -1) \  
$(x_1,x_2,x_3)\mapsto(x_1,-x_2,x_3)$.

\subsection{Proof of Lemma \ref{lemmadeltaijpsi}}\label{prooflemmadeltaijpsi}

First we recall that, denoting as $F(a,b;c;z)$ the Gauss hypergeometric function and as $(z)_n$ the Pochhammer's symbol, then, by definition,
\be\label{PochGauss}
(z)_n:=\frac{\Gamma(z+n)}{\Gamma(z)}\quad\mbox{and}\quad F(-n,b;c;z):=\sum_{m=0}^n{n\choose m} \frac{(-1)^m z^m (b)_m}{(c)_m}.
\ee
According to \cite{Stegun} p. 561 eq 15.4.6, one has
\be\label{GaussJaco}
F(-n,\alpha+1+\beta+n;\alpha+1;x)=\frac{n!}{(\alpha+1)_n}P_n^{(\alpha,\beta)}(1-2x),
\ee
where $P_n^{(\alpha,\beta)}$ is the Jacobi polynomial. \
From p. 556 eq. 15.1.1 one has
\be\label{simm}
F(a,b;c;z)=F(b,a;c;z),
\ee
p. 559  eq. 15.3.3
\be\label{abcGauss}
F(a,b;c;z)=(1-z)^{c-a-b}F(c-a,c-b;c;z)
\ee
and from p. 774
\be\label{normJaco}
\int_{-1}^1(1-x)^{\alpha}(1+x)^{\beta}P_n^{(\alpha,\beta)}(x)P_m^{(\alpha,\beta)}(x)dx=\frac{2^{\alpha+\beta+1}}{2n+\alpha+\beta+1}\frac{\Gamma(n+\alpha+1)\Gamma(n+\beta+1)}{n!\Gamma(n+\alpha+\beta+1)}\delta_{nm}.
\ee
In addition, we need the following
\begin{propo}
Let $l\geq s\geq h\geq -l$ and
\be \label{def_f}
f(l,h,s):=\left\{\begin{array}{ccc}
\prod_{j=h}^{s-1}\left[l(l+1)-j(j+1)\right]&\mbox{if}&h<s,\\
1&\mbox{if}& h=s;
\end{array}\right.
\ee
then
\be\label{Prodj(j+1)}
f(l,h,s)=\frac{(l-h)!(l+s)!}{(l+h)!(l-s)!}.
\ee
\proof
When $h=s$,
$$
f(l,h,h)=1=\frac{(l-h)!(l+h)!}{(l+h)!(l-h)!};
$$
assume that $h<s$ and (induction hypothesis)
$$
f(l,h,s-1)=\frac{(l-h)!(l+s-1)!}{(l+h)!(l-s+1)!},
$$
so
$$
f(l,h,s)=f(l,h,s-1)\left[l(l+1)-(s-1)s\right]=\frac{(l-h)!(l+s-1)!}{(l+h)!(l-s+1)!}(l+s)(l-s+1)=\frac{(l-h)!(l+s)!}{(l+h)!(l-s)!}.
$$
\endproof
\end{propo}
In the same way one can prove that, when $l\geq s\geq h\geq -l$, and setting
\be\label{def_g}
g(l,h,s):=\left\{\begin{array}{ccc}
\prod_{j=h+1}^{s}\left[l(l+1)-j(j-1)\right]&\mbox{if}&h<s,\\
1&\mbox{if}& h=s;
\end{array}\right.
\ee
then
\be\label{Prodj(j-1)}
g(l,h,s)=\frac{(l-h)!(l+s)!}{(l+h)!(l-s)!};
\ee
so, when $l\geq s\geq h\geq -l$,
\be \label{scambiohs}
\begin{split}
f(l,h,h)=1&=g(l,-h,-h)\quad\mbox{and}\\
f(l,h,s)=\prod_{j=h}^{s-1}\left[l(l+1)-j(j+1)\right]&=\prod_{j=-s+1}^{-h}\left[l(l+1)-j(j-1)\right]=g(l,-s,-h).
\end{split}
\ee
We need to point out that, when $0\leq n\leq h\leq l$,
\begin{equation*}
\begin{split}
A:=& \left\langle e^{2\log{\left(\cos{\frac{\theta}{2}} \right)}L_0} e^{{\tan{\frac{\theta}{2}}}L_+}\bpsi_l^h, e^{-{\tan{\frac{\theta}{2}}}L_+}\bpsi_l^n\right\rangle\\
\overset{(\ref{def_f})}=&\left\langle\sum_{s=h}^l\left(\cos{\frac{\theta}{2}}\right)^{2s} \frac{\left(\tan{\frac{\theta}{2}}\right)^{s-h}}{(s-h)!}\sqrt{f(l,h,s)}\bpsi_l^s,
\sum_{r=n}^l \frac{\left(-1\right)^{r-n}\left(\tan{\frac{\theta}{2}}\right)^{r-n}}{(r-n)!}\sqrt{f(l,n,r)}\bpsi_l^r \right\rangle\\
\overset{n\leq h}=& \left(-1\right)^{h-n} \sum_{s=h}^l (-1)^{s-h}\frac{\left[\tan{\frac{\theta}{2}} \right]^{s-h}}{(s-h)!}\sqrt{f(l,h,s)}\left(\cos{\frac{\theta}{2}}\right)^{2s}
\frac{\left[\tan{\frac{\theta}{2}} \right]^{s-n}}{(s-n)!}\sqrt{f(l,n,s)}\\
=& \left(-1\right)^{h-n}\sum_{s=h}^l (-1)^{s-h}\frac{1}{(s-h)!}\sqrt{f(l,h,s)}\left(\cos{\frac{\theta}{2}}\right)^{n+h}\left(\sin{\frac{\theta}{2}} \right)^{2s-n-h}
\frac{1}{(s-n)!}\sqrt{f(l,n,s)}\\
\overset{(\ref{Prodj(j+1)})}=& \left(-1\right)^{h-n}\left(\cos{\frac{\theta}{2}}\right)^{n+h}\left(\sin{\frac{\theta}{2}}\right)^{h-n}\sqrt{\frac{(l-n)!(l-h)!}{(l+n)!(l+h)!}}\sum_{s=h}^{l}(-1)^{s-h}\frac{(l+s)!}{(l-s)!(s-n)!(s-h)!}\left(\sin{\frac{\theta}{2}}\right)^{2(s-h)}\\
\overset{j=s-h}=& \left(-1\right)^{h-n}\left(\cos{\frac{\theta}{2}}\right)^{n+h}\left(\sin{\frac{\theta}{2}}\right)^{h-n}\sqrt{\frac{(l-n)!(l-h)!}{(l+n)!(l+h)!}}\sum_{j=0}^{l-h}(-1)^{j}\frac{(l+h+j)!}{(l-h-j)!(h-n+j)!(j)!}\left(\sin{\frac{\theta}{2}}\right)^{2j},\\
\end{split}
\end{equation*}
\bea
A &\overset{(\ref{PochGauss})}=& \left(-1\right)^{h-n}\left(\cos{\frac{\theta}{2}}\right)^{n+h}\left(\sin{\frac{\theta}{2}}\right)^{h-n}\sqrt{\frac{(l-n)!(l-h)!}{(l+n)!(l+h)!}}\nn
&&\cdot \frac{(l+h)!}{(l-h)!(h-n)!}F\left(-(l-h),l+h+1;h-n+1;\left(\sin{\frac{\theta}{2}}\right)^{2} \right)\nn
&\overset{(\ref{GaussJaco})}=& \left(-1\right)^{h-n}\left(\cos{\frac{\theta}{2}}\right)^{n+h}\left(\sin{\frac{\theta}{2}}\right)^{h-n}\sqrt{\frac{(l\!-\!n)!(l\!+\!h)!}{(l\!+\!n)!(l\!-\!h)!}}\,  \frac{(l\!-\!h)!(h\!-\!n)!}{(h\!-\!n)!(l\!-\!n)!} \, P_{l-h}^{(h-n,h+n)}\left(1-2\sin^2\frac{\theta}{2}\right)\nn
&=& \left(-1\right)^{h-n}\left(\cos{\frac{\theta}{2}}\right)^{n+h}\left(\sin{\frac{\theta}{2}}\right)^{h-n}\sqrt{\frac{(l\!-\!h)!(l\!+\!h)!}{(l\!+\!n)!(l\!-\!n)!}} \,  P_{l-h}^{(h-n,h+n)}\left(1-2\sin^2\frac{\theta}{2}\right),\label{e^-0+scalprod1}
\eea
\be 
\begin{split}
&\left\langle e^{-2\log{\left(\cos{\frac{\theta}{2}} \right)}L_0} e^{-{\tan{\frac{\theta}{2}}}L_-}\bpsi_l^{-h}, e^{{\tan{\frac{\theta}{2}}}L_-}\bpsi_l^{-n}\right\rangle\\
\overset{(\ref{def_g})}=&\left\langle\sum_{s=-h}^{-l}\left(\cos{\frac{\theta}{2}}\right)^{-2s} \frac{\left(-1\right)^{-s-h}\left(\tan{\frac{\theta}{2}}\right)^{-s-h}}{(-s-h)!}\sqrt{g(l,s,-h)}\bpsi_l^s,
\sum_{r=-n}^{-l} \frac{\left(\tan{\frac{\theta}{2}}\right)^{-r-n}}{(-r-n)!}\sqrt{g(l,r,-n)}\bpsi_l^r \right\rangle\\
\overset{(\ref{scambiohs})}=
&\left\langle\sum_{s=-h}^{-l}\left(\cos{\frac{\theta}{2}}\right)^{-2s} \frac{\left(-1\right)^{-s-h}\left(\tan{\frac{\theta}{2}}\right)^{-s-h}}{(-s-h)!}\sqrt{f(l,h,-s)}\bpsi_l^s,\sum_{r=-n}^{-l} \frac{\left(\tan{\frac{\theta}{2}}\right)^{-r-n}}{(-r-n)!}\sqrt{f(l,n,-r)}\bpsi_l^r \right\rangle\\
\overset{r, s\rightarrow -r,-s}=&\left\langle\sum_{s=h}^{l}\left(\cos{\frac{\theta}{2}}\right)^{2s} \frac{\left(-1\right)^{s-h}\left(\tan{\frac{\theta}{2}}\right)^{s-h}}{(s-h)!}\sqrt{f(l,h,s)}\bpsi_l^{-s},\sum_{r=n}^{l} \frac{\left(\tan{\frac{\theta}{2}}\right)^{r-n}}{(r-n)!}\sqrt{f(l,n,r)}\bpsi_l^{-r} \right\rangle\\
\overset{n\leq h}=&\sum_{s=h}^l (-1)^{s-h}\frac{\left[\tan{\frac{\theta}{2}} \right]^{s-h}}{(s-h)!}\sqrt{f(l,h,s)}\left(\cos{\frac{\theta}{2}}\right)^{2s}
\frac{\left[\tan{\frac{\theta}{2}} \right]^{s-n}}{(s-n)!}\sqrt{f(l,n,s)}
\end{split}\nonumber
\ee
\be 
\begin{split}
=&\sum_{s=h}^l (-1)^{s-h}\frac{\left[\tan{\frac{\theta}{2}} \right]^{s-h}}{(s-h)!}\sqrt{f(l,h,s)}\left(\cos{\frac{\theta}{2}}\right)^{2s}
\frac{\left[\tan{\frac{\theta}{2}} \right]^{s-n}}{(s-n)!}\sqrt{f(l,n,s)}\\
\overset{(\ref{e^-0+scalprod1})}=&\left(\cos{\frac{\theta}{2}}\right)^{n+h}\left(\sin{\frac{\theta}{2}}\right)^{h-n}\sqrt{\frac{(l-h)!(l+h)!}{(l+n)!(l-n)!}} P_{l-h}^{(h-n,h+n)}\left(1-2\sin^2\frac{\theta}{2}\right),
\end{split}\label{e^+0-scalprod1}
\ee
\begin{equation*}
\begin{split}
B:=& \left\langle e^{2\log{\left(\cos{\frac{\theta}{2}} \right)}L_0} e^{{\tan{\frac{\theta}{2}}}L_+}\bpsi_l^{-h}, e^{-{\tan{\frac{\theta}{2}}}L_+}\bpsi_l^n\right\rangle\\
\overset{(\ref{def_f})}=&\left\langle\sum_{s=-h}^l\left(\cos{\frac{\theta}{2}}\right)^{2s} \frac{\left(\tan{\frac{\theta}{2}}\right)^{s+h}}{(s+h)!}\sqrt{f(l,-h,s)}\bpsi_l^s,\sum_{r=n}^l \frac{\left(-1\right)^{r-n}\left(\tan{\frac{\theta}{2}}\right)^{r-n}}{(r-n)!}\sqrt{f(l,n,r)}\bpsi_l^r \right\rangle\\
\overset{-h\leq n}=&\sum_{s=n}^l (-1)^{s-n}\frac{\left[\tan{\frac{\theta}{2}} \right]^{s+h}}{(s+h)!}\sqrt{f(l,-h,s)}\left(\cos{\frac{\theta}{2}}\right)^{2s}
\frac{\left[\tan{\frac{\theta}{2}} \right]^{s-n}}{(s-n)!}\sqrt{f(l,n,s)}\\
=&\sum_{s=n}^l (-1)^{s-n}\frac{1}{(s+h)!}\sqrt{f(l,-h,s)}\left(\cos{\frac{\theta}{2}}\right)^{n-h}\left(\sin{\frac{\theta}{2}} \right)^{2s-n+h}
\frac{1}{(s-n)!}\sqrt{f(l,n,s)},
\end{split}
\end{equation*}
\begin{equation}
\begin{split}
B\overset{(\ref{Prodj(j+1)})}=&\left(\cos{\frac{\theta}{2}}\right)^{n-h}\left(\sin{\frac{\theta}{2}}\right)^{h+n}\sqrt{\frac{(l-n)!(l+h)!}{(l+n)!(l-h)!}}\sum_{s=n}^{l}(-1)^{s-n}\frac{(l+s)!}{(l-s)!(s-n)!(s+h)!}\left(\sin{\frac{\theta}{2}}\right)^{2(s-n)}\\
\overset{j=s-n}=&\left(\cos{\frac{\theta}{2}}\right)^{n-h}\left(\sin{\frac{\theta}{2}}\right)^{h+n}\sqrt{\frac{(l-n)!(l+h)!}{(l+n)!(l-h)!}}\\
&\cdot \sum_{j=0}^{l-n}(-1)^{j}\frac{(l+n+j)!}{(l-n-j)!(j)!(h+n+j)!}\left(\sin{\frac{\theta}{2}}\right)^{2j}\\
\overset{(\ref{PochGauss})}=&\left(\cos{\frac{\theta}{2}}\right)^{n-h}\left(\sin{\frac{\theta}{2}}\right)^{h+n}\sqrt{\frac{(l-n)!(l+h)!}{(l+n)!(l-h)!}}\\
&\cdot \frac{(l+n)!}{(l-n)!(h+n)!}F\left(-(l-n),l+n+1;h+n+1;\left(\sin{\frac{\theta}{2}}\right)^{2} \right)\\
\overset{(\ref{abcGauss})}=&\left(\cos{\frac{\theta}{2}}\right)^{n-h}\left(\sin{\frac{\theta}{2}}\right)^{h+n}\sqrt{\frac{(l+n)!(l+h)!}{(l-n)!(l-h)!}}\frac{1}{(h+n)!}\\
&\cdot \left[1-\left(\sin{\frac{\theta}{2}}\right)^{2}\right]^{h-n}F\left(l+h+1,-(l-h);h+n+1;\left(\sin{\frac{\theta}{2}}\right)^{2} \right) \\
\overset{(\ref{GaussJaco})}=&\left(\cos{\frac{\theta}{2}}\right)^{h-n}\left(\sin{\frac{\theta}{2}}\right)^{h+n}\sqrt{\frac{(l+n)!(l+h)!}{(l-n)!(l-h)!}}\frac{1}{(h+n)!}\\
&\cdot \frac{(l-h)!(h+n)!}{(l+n)!} P_{l-h}^{(h+n,h-n)}\left(1-2\sin^2{\frac{\theta}{2}} \right)\\
=&\left(\cos{\frac{\theta}{2}}\right)^{h-n}\left(\sin{\frac{\theta}{2}}\right)^{h+n}\sqrt{\frac{(l-h)!(l+h)!}{(l+n)!(l-n)!}} P_{l-h}^{(h+n,h-n)}\left(1-2\sin^2{\frac{\theta}{2}} \right)
\end{split}\label{e^-0+scalprod2}
\ee
and
\begin{equation*}
\begin{split}
D:=& \left\langle e^{-2\log{\left(\cos{\frac{\theta}{2}} \right)}L_0} e^{-{\tan{\frac{\theta}{2}}}L_-}\bpsi_l^{h}, e^{{\tan{\frac{\theta}{2}}}L_-}\bpsi_l^{-n}\right\rangle\\
\overset{(\ref{def_g})}=&\left\langle\sum_{s=h}^{-l}\left(\cos{\frac{\theta}{2}}\right)^{-2s} \frac{\left(-1\right)^{h-s}\left(\tan{\frac{\theta}{2}}\right)^{h-s}}{(h-s)!}\sqrt{g(l,s,h)}\bpsi_l^s,\sum_{s=-n}^{-l} \frac{\left(\tan{\frac{\theta}{2}}\right)^{-s-n}}{(-s-n)!}\sqrt{g(l,s,-n)}\bpsi_l^s \right\rangle\\
\overset{(\ref{scambiohs})}=&
\left\langle\sum_{s=h}^{-l}\left(\cos{\frac{\theta}{2}}\right)^{-2s} \frac{\left(-1\right)^{h-s}\left(\tan{\frac{\theta}{2}}\right)^{h-s}}{(h-s)!}\sqrt{f(l,-h,-s)}\bpsi_l^s,\sum_{s=-n}^{-l} \frac{\left(\tan{\frac{\theta}{2}}\right)^{-s-n}}{(-s-n)!}\sqrt{f(l,n,-s)}\bpsi_l^s \right\rangle\\
\overset{s\rightarrow-s}=&\left\langle\sum_{s=-h}^{l}\left(\cos{\frac{\theta}{2}}\right)^{2s} \frac{\left(-1\right)^{s+h}\left(\tan{\frac{\theta}{2}}\right)^{s+h}}{(s+h)!}\sqrt{f(l,-h,s)}\bpsi_l^{-s},\sum_{s=n}^{l} \frac{\left(\tan{\frac{\theta}{2}}\right)^{s-n}}{(s-n)!}\sqrt{f(l,n,s)}\bpsi_l^{-s} \right\rangle\\
\overset{-h\leq n}=&\sum_{s=n}^l (-1)^{s+h}\frac{\left[\tan{\frac{\theta}{2}} \right]^{s+h}}{(s+h)!}\sqrt{f(l,-h,s)}\left(\cos{\frac{\theta}{2}}\right)^{2s}
\frac{\left[\tan{\frac{\theta}{2}} \right]^{s-n}}{(s-n)!}\sqrt{f(l,n,s)},
\end{split}
\end{equation*}
\be
\begin{split}
D=&(-1)^{h+n}\sum_{s=n}^l (-1)^{s-n}\frac{\left[\tan{\frac{\theta}{2}} \right]^{s+h}}{(s+h)!}\sqrt{f(l,-h,s)}\left(\cos{\frac{\theta}{2}}\right)^{2s}
\frac{\left[\tan{\frac{\theta}{2}} \right]^{s-n}}{(s-n)!}\sqrt{f(l,n,s)}\\
\overset{(\ref{e^-0+scalprod2})}=& (-1)^{h+n}\left(\cos{\frac{\theta}{2}}\right)^{h-n}\left(\sin{\frac{\theta}{2}}\right)^{h+n}\sqrt{\frac{(l-h)!(l+h)!}{(l+n)!(l-n)!}} P_{l-h}^{(h+n,h-n)}\left(1-2\sin^2{\frac{\theta}{2}} \right).
\label{e^+0-scalprod2}
\end{split}
\ee
Finally, when $l\geq h\geq n\geq0$, one has
\be \label{contofattoriali1}
\begin{split}
\frac{(l-h)!(l+h)!}{(l+n)!(l-n)!}\frac{1}{2^{2h}}
\cdot \frac{2^{(h+n)+(h-n)+1}}{2(l-h)+(h+n)+(h-n)+1}&\\
\cdot \frac{\Gamma((l-h)+(h+n)+1)\Gamma((l-h)+(h-n)+1)}{(l-h)!\Gamma((l-h)+(h+n)+(h-n)+1)}&=\frac{2}{2l+1},
\end{split}
\ee
\be\label{contofattoriali2}
\begin{split}
\frac{(l-h)!(l+h)!}{(l+n)!(l-n)!}\frac{1}{2^{2h}}
\cdot \frac{2^{(h-n)+(h+n)+1}}{2(l-h)+(h-n)+(h+n)+1}&\\
\cdot \frac{\Gamma((l-h)+(h-n)+1)\Gamma((l-h)+(h+n)+1)}{(l-h)!\Gamma((l-h)+(h-n)+(h+n)+1)}&=\frac{2}{2l+1}.
\end{split}
\ee

We are now ready to prove the aforementioned lemma.

Assume that $0\leq n\leq h\leq l$; by means of the Gauss decomposition, 
$e^{i\theta L_2}$ can be written in the ``antinormal form'' 
(see e.g. eq. (4.3.14) in \cite{Perelomov})
\be\label{Gaussantinorm}
e^{i\theta L_2}
=e^{-{\tan{\frac{\theta}{2}}}L_-}e^{2\log{\left(\cos{\frac{\theta}{2}} \right)}L_0} e^{{\tan{\frac{\theta}{2}}}L_+};
\ee
hence
\bea
\begin{split}
& \int^{\pi}_0\!\!\!\!d\theta\sin\theta \:
\langle\bpsi_{j}^n,e^{i\theta L_2}\bpsi_{j}^{h}\rangle\: \langle e^{i\theta L_2}\bpsi_l^{h},\bpsi_l^n\rangle\\
=&\int^{\pi}_0\!\!\!\!d\theta\sin\theta \: \overline{\left\langle e^{2\log{\left(\cos{\frac{\theta}{2}} \right)}L_0} e^{{\tan{\frac{\theta}{2}}}L_+}\bpsi_j^h, e^{-{\tan{\frac{\theta}{2}}}L_+}\bpsi_j^n\right\rangle}
\left\langle e^{2\log{\left(\cos{\frac{\theta}{2}} \right)}L_0} e^{{\tan{\frac{\theta}{2}}}L_+}\bpsi_l^h, e^{-{\tan{\frac{\theta}{2}}}L_+}\bpsi_l^n\right\rangle\\
\overset{(\ref{e^-0+scalprod1})}=&2 (-1)^{2(h-n)} \int_0^{\pi}d\theta\left(\cos{\frac{\theta}{2}}\right)^{2(n+h)+1}\left(\sin{\frac{\theta}{2}}\right)^{2(h-n)+1}\sqrt{\frac{(l-h)!(l+h)!}{(l+n)!(l-n)!}}\sqrt{\frac{(j-h)!(j+h)!}{(j+n)!(j-n)!}} \\
&\cdot P_{l-h}^{(h-n,h+n)}\left(1-2\sin^2\frac{\theta}{2}\right)P_{j-h}^{(h-n,h+n)}\left(1-2\sin^2\frac{\theta}{2}\right)\\
\overset{x=1-2\sin^2\frac{\theta}{2}}=&\sqrt{\!\frac{(l\!-\!h)!(l\!+\!h)!}{(l\!+\!n)!(l\!-\!n)!}}\sqrt{\!\frac{(j\!-\!h)!(j\!+\!h)!}{(j\!+\!n)!(j\!-\!n)!}}
 \int\limits_{-1}^1\!\! \frac{dx}{2^{2h}}\, (1\!-\!x)^{h-n} (1\!+\!x)^{h+n}P_{l-h}^{(h-n,h+n)}\!\left(x\right)P_{j-h}^{(h-n,h+n)}\!\left(x\right)\\
\overset{(\ref{normJaco})\&(\ref{contofattoriali1})}=&\frac{2}{2l+1}\delta_{lj}.
\end{split}\nonumber
\eea
On the other hand, in order to calculate
$
\displaystyle\int^{\pi}_0\!\!\!\!d\theta\sin\theta \:
\langle\bpsi_{j}^{-n},e^{i\theta L_2}\bpsi_{j}^{-h}\rangle\: \langle e^{i\theta L_2}\bpsi_l^{-h},\bpsi_l^{-n}\rangle,
$
we can use now the ``normal form'' of the Gauss decomposition 
(see e.g. eq. (4.3.12) in \cite{Perelomov})
\be\label{Gaussnorm}
e^{i\theta L_2}
=e^{{\tan{\frac{\theta}{2}}}L_+}e^{-2\log{\left(\cos{\frac{\theta}{2}} \right)}L_0} e^{-{\tan{\frac{\theta}{2}}}L_-},
\ee
and we obtain
\bea
\begin{split}
& \int^{\pi}_0\!\!\!\!d\theta\sin\theta \:
\langle\bpsi_{j}^{-n},e^{i\theta L_2}\bpsi_{j}^{-h}\rangle\: \langle e^{i\theta L_2}\bpsi_l^{-h},\bpsi_l^{-n}\rangle\\
=&\int^{\pi}_0\!\!\!\!d\theta\sin\theta \: \overline{\left\langle e^{-2\log{\left(\cos{\frac{\theta}{2}} \right)}L_0} e^{-{\tan{\frac{\theta}{2}}}L_-}\bpsi_j^{-h}, e^{{\tan{\frac{\theta}{2}}}L_-}\bpsi_j^{-n}\right\rangle}
\left\langle e^{-2\log{\left(\cos{\frac{\theta}{2}} \right)}L_0} e^{-{\tan{\frac{\theta}{2}}}L_-}\bpsi_l^{-h}, e^{{\tan{\frac{\theta}{2}}}L_-}\bpsi_l^{-n}\right\rangle\\
\overset{(\ref{e^+0-scalprod1})}=&2 \int_0^{\pi}d\theta\left(\cos{\frac{\theta}{2}}\right)^{2(n+h)+1}\left(\sin{\frac{\theta}{2}}\right)^{2(h-n)+1}\sqrt{\frac{(l-h)!(l+h)!}{(l+n)!(l-n)!}}\sqrt{\frac{(j-h)!(j+h)!}{(j+n)!(j-n)!}} \\
&\cdot P_{l-h}^{(h-n,h+n)}\left(1-2\sin^2\frac{\theta}{2}\right)P_{j-h}^{(h-n,h+n)}\left(1-2\sin^2\frac{\theta}{2}\right)\\
\overset{x=1-2\sin^2\frac{\theta}{2}}=&\sqrt{\frac{(l\!-\!h)!(l\!+\!h)!}{(l\!+\!n)!(l\!-\!n)!}}\sqrt{\frac{(j\!-\!h)!(j\!+\!h)!}{(j\!+\!n)!(j\!-\!n)!}}\frac{1}{2^{2h}} 
\int\limits_{-1}^1\!\! dx \,(1\!-\!x)^{h-n} (1\!+\!x)^{h+n}P_{l-h}^{(h-n,h+n)}\!\left(x\right)P_{j-h}^{(h-n,h+n)}\!\left(x\right)\\
\overset{(\ref{normJaco})\& (\ref{contofattoriali1})}=&\frac{2}{2l+1}\delta_{lj}.
\end{split} \nonumber
\eea
Furthermore
\bea
\begin{split}
& \int^{\pi}_0\!\!\!\!d\theta\sin\theta \:
\langle\bpsi_{j}^{n},e^{i\theta L_2}\bpsi_{j}^{-h}\rangle\: \langle e^{i\theta L_2}\bpsi_l^{-h},\bpsi_l^{n}\rangle\\
\overset{(\ref{Gaussantinorm})}=&\int^{\pi}_0\!\!\!\!d\theta\sin\theta \: \overline{\left\langle e^{2\log{\left(\cos{\frac{\theta}{2}} \right)}L_0} e^{{\tan{\frac{\theta}{2}}}L_+}\bpsi_j^{-h}, e^{-{\tan{\frac{\theta}{2}}}L_+}\bpsi_j^n\right\rangle}
\left\langle e^{2\log{\left(\cos{\frac{\theta}{2}} \right)}L_0} e^{{\tan{\frac{\theta}{2}}}L_+}\bpsi_l^{-h}, e^{-{\tan{\frac{\theta}{2}}}L_+}\bpsi_l^n\right\rangle\\
\overset{(\ref{e^-0+scalprod2})}=&2 \int_0^{\pi}d\theta\left(\cos{\frac{\theta}{2}}\right)^{2(h-n)+1}\left(\sin{\frac{\theta}{2}}\right)^{2(h+n)+1}\sqrt{\frac{(l-h)!(l+h)!}{(l+n)!(l-n)!}}\sqrt{\frac{(j-h)!(j+h)!}{(j+n)!(j-n)!}} \\
&\cdot P_{l-h}^{(h+n,h-n)}\left(1-2\sin^2\frac{\theta}{2}\right)P_{j-h}^{(h+n,h-n)}\left(1-2\sin^2\frac{\theta}{2}\right)\\
\overset{x=1-2\sin^2\frac{\theta}{2}}=&\sqrt{\frac{(l\!-\!h)!(l\!+\!h)!}{(l\!+\!n)!(l\!-\!n)!}}\sqrt{\frac{(j\!-\!h)!(j\!+\!h)!}{(j\!+\!n)!(j\!-\!n)!}}\frac{1}{2^{2h}} 
\int\limits_{-1}^1\!\! dx\, (1\!-\!x)^{h+n} (1\!+\!x)^{h-n}P_{l-h}^{(h+n,h-n)}\!\left(x\right)P_{j-h}^{(h+n,h-n)}\!\left(x\right)\\
\overset{(\ref{normJaco})\&(\ref{contofattoriali2})}=&\frac{2}{2l+1}\delta_{lj}
\end{split}\nonumber
\eea
and, finally, as claimed,
\begin{equation*}
\begin{split}
E:=& \int^{\pi}_0\!\!\!\!d\theta\sin\theta \:
\langle\bpsi_{j}^{-n},e^{i\theta L_2}\bpsi_{j}^{h}\rangle\: \langle e^{i\theta L_2}\bpsi_l^{h},\bpsi_l^{-n}\rangle\\
\overset{(\ref{Gaussnorm})}=&\int^{\pi}_0\!\!\!\!d\theta\sin\theta \: \overline{\left\langle e^{-2\log{\left(\cos{\frac{\theta}{2}} \right)}L_0} e^{-{\tan{\frac{\theta}{2}}}L_-}\bpsi_j^{h}, e^{{\tan{\frac{\theta}{2}}}L_-}\bpsi_j^{-n}\right\rangle}
\left\langle e^{-2\log{\left(\cos{\frac{\theta}{2}} \right)}L_0} e^{-{\tan{\frac{\theta}{2}}}L_-}\bpsi_l^{h}, e^{{\tan{\frac{\theta}{2}}}L_-}\bpsi_l^{-n}\right\rangle\\
\\
\overset{(\ref{e^+0-scalprod2})}=&2 (-1)^{2(h+n)} \int_0^{\pi}d\theta\left(\cos{\frac{\theta}{2}}\right)^{2(h-n)+1}\left(\sin{\frac{\theta}{2}}\right)^{2(h+n)+1}\sqrt{\frac{(l-h)!(l+h)!}{(l+n)!(l-n)!}}\sqrt{\frac{(j-h)!(j+h)!}{(j+n)!(j-n)!}} \\
&\cdot P_{l-h}^{(h+n,h-n)}\left(1-2\sin^2\frac{\theta}{2}\right)P_{j-h}^{(h+n,h-n)}\left(1-2\sin^2\frac{\theta}{2}\right),
\end{split}
\end{equation*}
\begin{equation*}
\begin{split}
E\overset{x=1-2\sin^2\frac{\theta}{2}}=&\sqrt{\frac{(l\!-\!h)!(l\!+\!h)!}{(l\!+\!n)!(l\!-\!n)!}}\sqrt{\frac{(j\!-\!h)!(j\!+\!h)!}{(j\!+\!n)!(j\!-\!n)!}}\frac{1}{2^{2h}}  \int\limits_{-1}^1\!\! dx\, (1\!-\!x)^{h+n} (1\!+\!x)^{h-n}P_{l-h}^{(h+n,h-n)}\!\left(x\right)P_{j-h}^{(h+n,h-n)}\!\left(x\right)\\
\overset{(\ref{normJaco})\&(\ref{contofattoriali2})}=&\frac{2}{2l+1}\delta_{lj}.
\end{split}
\end{equation*}

\subsection{Proofs of some results regarding $S^2_\Lambda$}
\label{Ciccio}

{\bf Proof  of (\ref{LXURomega})}. $L_+\bomega^\beta=0$, $L_-\bomega^\beta$ is a combination
of $\bpsi_l^{l-1}$, therefore is orthogonal to $\bomega^\beta$. Hence
$$
\langle L_\pm\rangle_{\bomega^\beta}\!=\!0,\quad\Rightarrow\quad   |\langle \bL\rangle_{\bomega^\beta}|=\langle L_0\rangle_{\bomega^\beta}=\sum_{l=0}^{\Lambda}\frac{l(2l\!+\!1)}{(\Lambda\!+\!1)^2}\overset{(\ref{us_sumh_1})_2}=\frac{\Lambda(4\Lambda\!+\!5)}{6(\Lambda\!+\!1)};
$$
while
$$
\left\langle \bm{L}^2\right\rangle_{\bm{\omega}^{\beta}}=\sum_{l=0}^{\Lambda}\frac{l(l+1)(2l+1)}{(\Lambda+1)^2}\overset{(\ref{us_sumh_2})}=\frac{\frac{1}{2}\Lambda(\Lambda+1)^2(\Lambda+2)}{(\Lambda+1)^2}=\frac{\Lambda(\Lambda+2)}{2}.
$$
Replacing these results in \ $(\Delta \bL)^2_{\bomega^\beta} =\langle\bL^2\rangle_{\bomega^\beta}-\langle \bL\rangle_{\bomega^\beta}^2$, we find
$$
(\Delta \bL)^2_{\bomega^{\beta}}=\frac{\Lambda(\Lambda+2)}{2}-\left(\frac{\Lambda(4\Lambda\!+\!5)}{6(\Lambda\!+\!1)}\right)^2=\frac{\Lambda(2\Lambda^3\!+\!32\Lambda^2\!+\!65\Lambda\!+\!36)}{36(\Lambda\!+\!1)^2}.
$$
On the other hand,  
$x_0\bomega^\beta$ is a combination of $\bpsi_{l-1}^{l}, \bpsi_{l+1}^{l}$, therefore is orthogonal to $\bomega^\beta$, and $\langle x_0\rangle=0$. Hence
$$
\left\langle \bx\right\rangle^2=\left\langle x_1\right\rangle^2+ \left\langle x_2\right\rangle^2+ \left\langle x_3\right\rangle^2=\frac{\left\langle x_+ +x_-   \right\rangle^2}{4}-\frac{\left\langle x_+-x_-\right\rangle^2}{4}+\left\langle x_0\right\rangle^2=\left\langle x_+\right\rangle\left\langle x_-\right\rangle=\left\vert\left\langle x_+\right\rangle\right\vert^2,
$$
$$
\left(\Delta\bx\right)^2=\left\langle\bx^2\right\rangle-\left\vert\left\langle x_+\right\rangle\right\vert^2.
$$
But
\bea
\left\langle\bm{x}^2\right\rangle_{\bm{\omega}^{\beta}}\overset{(\ref{xx})_2}= 1+\sum_{l=0}^{\Lambda}\frac{l(l+1)+1}{k}\frac{2l+1}{(\Lambda+1)^2}
-\left[1+\frac{(\Lambda\!+\!1)^2}{k}\right]\frac 1{\Lambda\!+\!1}\nn
=\frac {\Lambda}{\Lambda\!+\!1}-\frac{\Lambda\!+\!1}{k}
+\frac1{k(\Lambda+1)^2}\left[2\sum_{l=0}^{\Lambda}l(l+1)(l+2)-3\sum_{l=0}^{\Lambda}l(l+1)+\sum_{l=0}^{\Lambda}(2l+1)\right]\nn
\overset{(\ref{us_sumh_3})_1}=\!\frac {\Lambda}{\Lambda\!+\!1}\!-\!\frac{\Lambda\!+\!1}{k}\!+\!\frac1{k(\Lambda\!+\!1)^2}\!\left[\frac {\Lambda}2 (\Lambda\!+\!1)(\Lambda\!+\!2)(\Lambda\!+\!3)-\Lambda(\Lambda\!+\!1)(\Lambda\!+\!2)+(\Lambda+1)^2\right]\nn
=\frac {\Lambda}{\Lambda\!+\!1}-\frac{\Lambda\!+\!1}{k}
+\frac{1+(\Lambda+1)^2}{2k}
=\frac {\Lambda}{\Lambda\!+\!1}+\frac{\Lambda^2}{2k}
\overset{(\ref{kineq})}
\leq \frac {\Lambda}{\Lambda\!+\!1}+\frac{1}{2(\Lambda+1)^2},
\qquad\qquad \label{<x^2>}
\eea
while
\begin{equation*}
\begin{split}
x_+\bm{\omega}^{\beta}&\overset{(\ref{azionex})}=\sum_{l=0}^{\Lambda-1}e^{i\beta_l}\frac{\sqrt{2l+1}}{\Lambda+1}c_{l+1}B_l^{+,l}\bpsi_{l+1}^{l+1}\overset{(\ref {Clebsch})}=\sum_{l=0}^{\Lambda-1}e^{i\beta_l}\frac{\sqrt{2l+1}}{\Lambda+1}c_{l+1}\left(-\sqrt{\frac{2l+2}{2l+3}} \right)\bpsi_{l+1}^{l+1}\\
&=-\sum_{l=1}^{\Lambda}e^{i\beta_{l-1}}\sqrt{\frac{(2l)(2l-1)}{2l+1}}\frac{c_l}{\Lambda+1}\bpsi_l^l \quad\Longrightarrow\\
\left\langle x_+\right\rangle_{\bm{\omega}^{\beta}}&=-\sum_{l=1}^{\Lambda}e^{i\left(\beta_{l-1}-\beta_l \right)}\frac{c_l\sqrt{(2l)(2l-1)}}{(\Lambda+1)^2},
\end{split}
\end{equation*}
so
\begin{equation*}
\langle \bm{x}\rangle_{\bm{\omega}^{\beta}}^2=\left\vert\left\langle x_+\right\rangle_{\bm{\omega}^{\beta}}\right\vert^2=\left\vert\sum_{l=1}^{\Lambda}\frac{c_l2l\sqrt{1-\frac{1}{2l}}}{(\Lambda+1)^2}e^{i(\beta_{l-1}-\beta_l)}\right\vert^2.
\end{equation*}
Since all $\frac{c_l\sqrt{(2l)(2l-1)}}{(\Lambda+1)^2} > 0$, to maximize $\left\vert\langle x_+\rangle_{\bomega^\beta}\right\vert$,
and thus minimize $(\Delta \bx)^2_{\bomega^\beta} $, we need to take all the  $\beta_l$ equal (mod. $2\pi$), in particular $\beta_l=0$. 

In this case, if we use $\sqrt{1\!-\!\frac{1}{2l}}\geq 1-\frac{1}{2l}$ $\forall l\in\mathbb{N}$ and $c_l\geq 1$, we get (here and below $\bm{\omega}\equiv \bm{\omega}^0$)
\begin{equation*}
\begin{split}
\langle \bm{x}\rangle_{\bm{\omega}}^2& \geq \left[\frac{2}{(\Lambda+1)^2}\sum_{l=1}^{\Lambda}l\left(1-\frac{1}{2l}\right)\right]^2\overset{(\ref{us_sumh_3})_2}= \left\{\frac{2}{(\Lambda+1)^2}\left[\frac{\Lambda^2}{2}\right]\right\}^2=\frac{\Lambda^4}{(\Lambda+1)^4}.
\end{split}
\end{equation*}
Finally, we find
\begin{equation*}
\begin{split}
\left(\Delta\bm{x}\right)^2_{\bm{\omega}}&=\left\langle\bm{x}^2\right\rangle_{\bm{\omega}}-\left\langle\bm{x}\right\rangle_{\bm{\omega}}^2\le\frac {\Lambda}{\Lambda\!+\!1}+\frac{1}{2(\Lambda+1)^2}-\frac{\Lambda^4}{(\Lambda+1)^4} =\frac{2\Lambda(\Lambda+1)^3+(\Lambda+1)^2-2\Lambda^4}{2(\Lambda+1)^4}\\
&=\frac{6\Lambda^3+7\Lambda^2+4\Lambda+1}{2(\Lambda+1)^4}<\frac{3\Lambda^3+9\Lambda^2+9\Lambda+3}{(\Lambda+1)^4}=\frac{3}{\Lambda+1}.
\end{split}
\end{equation*}

\noindent
{\bf Proof of (\ref{LXURphi})}. $L_0\bphi^\beta\!=0$,
while $L_\pm\bphi^\beta$ are combinations
of $\bpsi_l^{\pm 1}$, therefore are orthogonal to $\bphi^\beta$; similarly, 
$x_{\pm}\bphi^\beta$ are combinations
of $\bpsi_{l-1}^{\pm 1}, \bpsi_{l+1}^{\pm 1}$, therefore are orthogonal to $\bphi^\beta$. Hence
$$
\langle L_a\rangle_{\bphi^\beta}=0,\:\:\:\langle x_\pm\rangle_{\bphi^\beta}=0\qquad \Rightarrow
\qquad \langle \bL\rangle_{\bphi^\beta}=0,\quad|\langle \bx\rangle_{\bphi^\beta}|=|\langle x_0\rangle_{\bphi^\beta}|.
$$
Replacing these results in \ $(\Delta \bL)^2=\langle\bL^2\rangle_{\bphi^\beta}-\langle \bL\rangle_{\bphi^\beta}^2$  \ and
using  (\ref{ResolIdS^2_Lphi}), we find, as claimed  
$$
(\Delta \bL)^2=\langle\bL^2\rangle_{\bphi^\beta}= \langle\bphi^\beta, \bL^2\bphi^\beta\rangle
=\sum_{l=1}^{\Lambda}\frac{l(l\!+\!1)(2l\!+\!1)}{(\Lambda\!+\!1)^2}\stackrel{(\ref{us_sumh_2})}{=}
\frac{\Lambda(\Lambda\!+\!2)}{2}.
$$
On the other hand,
\begin{equation*}
\begin{split}
x^0\bm{\phi}^{\beta}\overset{(\ref{azionex})}=&\sum_{l=0}^{\Lambda}e^{i\beta_l}\frac{\sqrt{2l+1}}{\Lambda+1}\left(c_lA_l^{0,0}\bpsi_{l-1}^0+c_{l+1}B_l^{0,0}\bpsi_{l+1}^0\right)\\
\overset{(\ref{Clebsch})}=&\sum_{l=0}^{\Lambda}\frac{e^{i\beta_l}}{\Lambda+1}\left(c_l\frac{l}{\sqrt{2l-1}}\bpsi_{l-1}^0+c_{l+1}\frac{l+1}{\sqrt{2l+3}}\bpsi_{l+1}^0   \right)\\
=&\sum_{l=1}^{\Lambda}\frac{e^{i\beta_l}c_l}{\Lambda+1}\frac{l}{\sqrt{2l-1}}\bpsi_{l-1}^0+\sum_{l=0}^{\Lambda-1}\frac{e^{i\beta_l}c_{l+1}}{\Lambda+1}\frac{l+1}{\sqrt{2l+3}}\bpsi_{l+1}^0\\
=&\sum_{l=0}^{\Lambda-1}\frac{e^{i\beta_{l+1}}c_{l+1}}{\Lambda+1}\frac{l+1}{\sqrt{2l+1}}\bpsi_l^0+\sum_{l=1}^{\Lambda}\frac{e^{i\beta_{l-1}}c_{l}}{\Lambda+1}\frac{l}{\sqrt{2l+1}}\bpsi_l^0,\quad\Longrightarrow\\
\left\langle x^0\right\rangle_{\bm{\phi}^{\beta}}=&\sum_{l=0}^{\Lambda-1}e^{i(\beta_{l+1}-\beta_l)}c_{l+1}\frac{l+1}{(\Lambda+1)^2}+\sum_{l=1}^{\Lambda}e^{i(\beta_{l-1}-\beta_l)}c_l\frac{l}{(\Lambda+1)^2}\\
=&\sum_{l=1}^{\Lambda}\frac{2lc_l}{(\Lambda+1)^2}\cos{(\beta_{l-1}-\beta_l)},
\end{split}
\end{equation*}
this means that $\left\langle x^0\right\rangle_{\bm{\phi}^{\beta}}^2\equiv \left\langle \bm{x}\right\rangle_{\bm{\phi}^{\beta}}^2 $ is maximal when $\beta\equiv0$, and in this case one has (here and on $\bm{\phi}\equiv\bm{\phi}^0$)
$$
\left\langle \bm{x}\right\rangle_{\bm{\phi}}^2\overset{c_l \geq 1} \geq \left[\sum_{l=1}^{\Lambda}\frac{2l}{(\Lambda+1)^2}\right]^2\overset{(\ref{sum_generic})}=\frac{\Lambda^2}{(\Lambda+1)^2}.
$$
One  easily checks that \ 
$\langle\bx^2\rangle_{\bphi^\beta}=\langle\bx^2\rangle_{\bomega^\beta}$; \ hence, using (\ref{<x^2>}), on $\bphi$ it follows, as claimed
\bea
(\Delta \bx)^2_{\bm{\phi}}=\langle\bx^2\rangle_{\bphi}-\langle \bx\rangle_{\bphi}^2\leq \frac {\Lambda}{\Lambda\!+\!1}+\frac{1}{2(\Lambda+1)^2}-\frac{\Lambda^2}{(\Lambda\!+\!1)^2}\nn
=\frac{2\Lambda(\Lambda+1)+1-2\Lambda^2}{2(\Lambda+1)^2}
=\frac{2\Lambda+1}{2(\Lambda+1)^2}<\frac{1}{\Lambda+1}.
\nonumber
\eea

\noindent
{\bf Proof of (\ref{uncfinalenostra})}. 
\be\label{xxtildechi}
\begin{split}
\left\langle \bx^2\right\rangle_{\widetilde{\bm{\chi}}}\overset{(\ref{xx})_2}=
&\sum_{l=0}^{\Lambda}\left\vert \widetilde{\chi}^l \right\vert^2+\frac{\left[\sum_{l=0}^{\Lambda}l(l+1)\left\vert \widetilde{\chi}^l \right\vert^2\right]+1}{k(\Lambda)}-\left[1+\frac{(\Lambda+1)^2}{k(\Lambda)}\right]\frac{\Lambda+1}{2\Lambda+1}\left\vert \widetilde{\chi}^{\Lambda} \right\vert^2 \\
\overset{\left\|\widetilde{\bm{\chi}}\right\|_2=1}=&1+\frac{\left[\sum_{l=0}^{\Lambda}l(l+1)\left\vert \widetilde{\chi}^l \right\vert^2\right]+1}{k(\Lambda)}-\left[1+\frac{(\Lambda+1)^2}{k(\Lambda)}\right]\frac{\Lambda+1}{2\Lambda+1}\left\vert \widetilde{\chi}^{\Lambda} \right\vert^2 \\
\leq &1+\frac{\frac{2}{\Lambda+2}\left[\sum_{l=1}^{\Lambda}l(l+1)\right]+1}{k(\Lambda)}\overset{(\ref{sum_generic})}\leq 1+\frac{\frac 23 \Lambda(\Lambda+1)+1}{k(\Lambda)} \overset{(\ref{kineq})}\leq 1+\frac{\frac 23 \Lambda(\Lambda+1)+1}{\Lambda^2(\Lambda+1)^2},
\end{split}
\ee
so, putting together (\ref{disscalprod}) and (\ref{xxtildechi}), we obtain, as claimed,
\bea
(\Delta \bm{x})^2_{\widetilde{\bm{\chi}}}
=\left\langle \bx^2\right\rangle_{\widetilde{\bm{\chi}}}-\left\langle \bx\right\rangle_{\widetilde{\bm{\chi}}}^2
=\left\langle \bx^2\right\rangle_{\widetilde{\bm{\chi}}}-\left\langle x_0\right\rangle_{\widetilde{\bm{\chi}}}^2
<1-\cos^2\left(\frac{\pi}{\Lambda\!+\!2}\right) +\frac{\frac 23 \Lambda(\Lambda+1)+1}{\Lambda^2(\Lambda+1)^2}\nn
=\sin^2\left(\frac{\pi}{\Lambda\!+\!2}\right)+\frac{\frac 23 +\frac 2{3\Lambda}+\frac 1{\Lambda^2}}{(\Lambda+1)^2} \overset{\Lambda\ge 3}<\frac{\pi^2}{(\Lambda+2)^2}+\frac{1}{(\Lambda+1)^2}<\frac{11}{(\Lambda+1)^2}.
\nonumber 
\eea

\end{document}